%% file: CKS1_arXiv_v1.tex
\documentclass[12pt]{article}
\pdfoutput=1 
\usepackage[utf8]{inputenc}

\usepackage{amsmath,amsfonts,amssymb,epsfig}
\usepackage{slashed}
\usepackage{cite}
\usepackage{multirow}
\usepackage{multicol}
\usepackage{cite}
\usepackage{bbm}

\usepackage{xcolor}
\usepackage[
   bookmarks=true,
   linktocpage=true,
   colorlinks=true,
   allbordercolors=white, allcolors=blue
]{hyperref}

\usepackage{calc}
\usepackage{rotating}
\usepackage[english]{babel}
\usepackage{pifont}

\usepackage{graphicx}
\usepackage{subfig}
\usepackage{float}
\usepackage{amsmath}
\usepackage{amssymb}
\usepackage{amsthm}
\usepackage{latexsym}
\usepackage{dcolumn}
\usepackage{geometry}
\usepackage{mciteplus}
\usepackage{fancyhdr}
\usepackage[normalem]{ulem}
\usepackage[utf8]{inputenc}
\usepackage{caption}
\usepackage{array}
\usepackage{sectsty}
\allsectionsfont{\boldmath}
\usepackage{bm}

\def\gtwid{\mathrel{\raise.3ex\hbox{$>$\kern-.75em\lower1ex\hbox{$\sim
$}}}}
\def\vio{\mathrel{\hbox{$E$\kern-.60em\hbox{$/
$}}}}
\newcolumntype{P}[1]{>{\centering\arraybackslash}p{#1}}

\input macros.tex

\restylefloat{figure}

\newcommand{\email}[1]{\href{mailto:#1}{#1}}
 
\begin{document}
\title{\LARGE {\bf Perturbative unitarity bounds from qubit entanglement in scalar scattering and QED}}
\author{\\ Shiuli Chatterjee\footnote{\email{shiuli.chatterjee@ncbj.gov.pl}},\,
Kamila Kowalska\footnote{\email{kamila.kowalska@ncbj.gov.pl}},\,
and Enrico Maria Sessolo\footnote{\email{enrico.sessolo@ncbj.gov.pl}}\\[2ex]
\small {\em National Centre for Nuclear Research}\\
\small {\em Pasteura 7, 02-093 Warsaw, Poland  }\\
}
%
\date{}
\maketitle
\thispagestyle{fancy}
\begin{abstract}
In the framework of $2\to 2$ relativistic scattering described by the perturbative $S$-matrix, we perform a comprehensive analysis of the properties of quantum channels obtained by tracing out the momentum from the scattered state. We work in two distinct scenarios with qubit degrees of freedom: flavored scalar contact interactions and QED, focusing in particular on initial-state leptons with transverse polarization. We identify the conditions that make the quantum channel unitary, calculate the entanglement of the qubits through the concurrence, and use its maximally allowed value to extract perturbativity upper bounds on the coupling constants. We find that the limits are often more constraining than the corresponding partial-wave expansion bounds, and that they allow us to bypass some of the obstacles associated in the latter with the singular behavior of forward scattering.      
\end{abstract}
\newpage 

\tableofcontents

\setcounter{footnote}{0}

\section{Introduction\label{sec:intro}}

There has been growing interest in establishing conceptual links between quantum information theory (QIT) and quantum filed theory, in particular in the context of relativistic scattering\cite{Seki:2014cgq,Peschanski:2016hgk,Carney:2016tcs,Faleiro:2016lsf,Cervera-Lierta:2017tdt,Peschanski:2019yah,Beane:2018oxh, Beane:2020wjl,Low:2021ufv, Beane:2021zvo, Fonseca:2021uhd,Liu:2022grf,Fedida:2022izl,Carena:2023vjc,Duch:2024pwm,Kowalska:2024kbs,Low:2024mrk,Low:2024hvn,Chang:2024wrx, Thaler:2024anb,Blasone:2024dud,Blasone:2024jzv, DuasoPueyo:2024usw,
Blasone:2025ddi,Blasone:2025tor,Garrido:2025xpk,Busoni:2025dns,McGinnis:2025brt,Hu:2025lua, Carena:2025wyh, Sou:2025tyf,Kowalska:2025qmf,Cheng:2025zaw,Guzman:2025tgt,Nunez:2025xds,Lyu:2025lja,McGinnis:2025iab,McGinnis:2025xgt,Liu:2025iwh,Pacheco:2025qam,Peschanski:2026edo,Liu:2026dzv,Fang:2026ddi,Xie:2026ixr,Low:2026evp,Lee:2026bvm,Fabbrichesi:2026jvm,Shivashankara:2026nzq,Li:2026yjs,Li:2026zum,Yamagishi:2026ihf,Kanemura:2026won,Jangid:2026uvf,Zheng:2026pma,Liu:2026gju,Chang:2026nzq,Chang:2026xif}. The possibility of testing QIT-based observables at high-energy colliders has also attracted attention\cite{ATLAS:2023fsd,CMS:2024pts,Afik:2020onf,Fabbrichesi:2021npl,Severi:2021cnj,Barr:2021zcp,Altakach:2022ywa,Afik:2022kwm,Aoude:2022imd,Aguilar-Saavedra:2022uye,Fabbrichesi:2022ovb,Severi:2022qjy,Barr:2022wyq,Aguilar-Saavedra:2022wam,Ashby-Pickering:2022umy,Dong:2023xiw,Varma:2023gwh,Aguilar-Saavedra:2023hss,Han:2023fci,Ma:2023yvd,Ehataht:2023zzt,Fabbrichesi:2023cev,Aoude:2023hxv,Bi:2023uop,Bernal:2023ruk,Altomonte:2024upf,Maltoni:2024tul,Aguilar-Saavedra:2024hwd,Barr:2024djo,Goncalves:2025mvl,Goncalves:2025xer}.  In particle physics a scattering event is described by the action of a unitary operator, the $S$-matrix, on a multi-particle state, which changes the kinematical properties, quantum numbers, type, and even overall number of the particles. This picture is to a great extent equivalent to the action of a quantum gate, which transforms an initial collection of qudits into a new set, whose information theoretical properties, like for example purity or entanglement, may be different. One might even go so far as to say that a scattering event is Nature’s own quantum computation.

The closed-system evolution of a quantum state must be unitary. In particle physics, the unitarity of the $S$-matrix is equivalent to the conservation of probability, which must sum up to one for all possible scattering outcomes including, importantly, the trivial one. 
In QIT, a fundamental implication of unitarity is that quantum gates preserve information. 
At the same time, nontrivial transformations of quantum correlations may, for example, generate entanglement between initially separable subsystems, or destroy entanglement that was already present.

To address the question of how a scattering event, acting as a quantum gate, distributes information over the entire quantum system, one needs to take a closer look at the degrees of freedom that the system has at its disposal. While any non-trivial interaction of two scattering particles induces correlations between different modes of their outgoing momenta, in a way that is proportional to the total cross section\cite{Balasubramanian:2011wt,Hsu:2012gk,Seki:2014cgq,Peschanski:2016hgk,Carney:2016tcs,Faleiro:2016lsf,Peschanski:2019yah}, 
quantum correlations between discrete quantum numbers (spin, polarization, flavor, etc.) typically reflect internal features of the $S$-matrix, and may even point to specific properties of the underlying Lagrangian. In order to access these qudit-qudit correlations, one can trace the momentum degrees of freedom out of the final-state density matrix.
In the language of QIT, such a procedure defines a quantum channel acting on the qudit degrees of freedom, i.e., 
a linear transformation that maps the initial density matrix onto the final state, while preserving its trace and positivity.\footnote{Instead of tracing out the entire momentum Hilbert space, one can project the final-state density matrix onto a fixed momentum. Such an operation, called a quantum instrument, is not trace preserving and requires a post-measurement normalization of the density matrix.} In Ref.\cite{Kowalska:2024kbs,Kowalska:2025qmf} some of us applied this approach to evaluate the amount of entanglement generated or lost by contact interactions in the two-Higgs doublet model~(2HDM) and in high-energy electron-positron annihilation processes in quantum electrodynamics~(QED). 

A special class of quantum channels are unitary channels, which have the additional property of  preserving the spectrum of a density matrix. 
As a result, a unitary channel allows one to encapsulate the evolution of a quantum subsystem in a closed-system dynamics, while still allowing nontrivial transformations of quantum correlations, entanglement, and quantum information.

In this work, we perform a comprehensive analysis of the properties of quantum channels that appear in the framework of $2\to 2$ relativistic scattering described by the perturbative $S$-matrix. We focus on two scenarios: first the simpler case of two flavor qubits interacting through contact interactions, which is the situation realized in the 2HDM at ultrarelativistic energies. Then, we take on high-energy scattering in QED, where we identify the qubits with the helicity of the scattered leptons. We consider longitudinal and transverse polarization of the initial state and we finally extend the analysis to include the flavor of the scattered leptons as an additional quantum degree of freedom. In each case, we identify the conditions that make the quantum channel, induced by tracing out the momentum, unitary. We find that whether or not such a unitary channel can be established at the lowest order in perturbation theory depends on the kinematic structure of the scattering amplitude and on the coherence of the initial state. 

We quantify the entanglement between the qubits of the final state by computing the concurrence. We find that requiring that the concurrence is physical imposes a perturbative upper bound on the strength of the interactions responsible for generating non-classical correlations between the qubit degrees of freedom. Note that, in a similar spirit, Ref.\cite{DuasoPueyo:2024usw} used the interaction-driven increase of entanglement in momentum space, captured by the purity of the density matrix, as a diagnostic for the breakdown of perturbation theory. Here we focus instead on the QIT properties of the two-qubit system after the momentum is traced out. For all processes considered, we find that bounds arising from the physicality of the concurrence are consistent with the corresponding constraints on partial-wave amplitudes. Interestingly, in the case of QED, we overcome some of the complications undermining the traditional partial-wave approach, which typically requires the introduction of a renormalization scheme to absorb the singularities associated with forward/backward scattering in the $t$ and $u$ channels.

The paper is organized as follows. In \refsec{sec:forma}, we briefly review the quantum features of $2\to 2$ scattering, closely following the formalism of Refs.\cite{Kowalska:2024kbs,Kowalska:2025qmf}, and we provide a short summary of the characteristics of quantum channels. In \refsec{sec:scalar}, we derive the perturbative unitarity bounds on the quartic couplings of the 2HDM, emerging from the entanglement of two flavor qubits. In \refsec{sec:QED}, we perform an analogous analysis for Moeller scattering, Bhabha scattering, and $e^- e^+$ annihilation in QED, considering both longitudinal and transverse polarization of the initial state, and the possibility of helicity and flavor qubits. We derive perturbative unitarity bounds in those cases, and confront them with the bounds obtained from a partial-wave expansion. We summarize our findings and conclude in \refsec{sec:conc}. Appendix~\ref{app:Amplitudes} presents the full expressions of the QED scattering amplitudes used in this work. Appendix~\ref{app:proj} presents a comparison of our QED results with the findings in the projection approach, which is based on quantum instruments rather than quantum channels.

\section{Quantum properties of $2\to 2$ scattering \label{sec:forma}}

We consider $2\to 2$ scattering in a perturbative quantum field theory. Our goal is to look at the QIT properties of a scattering event 
of two particles, each carrying a set of generalized ``quantum numbers.'' These include, first and foremost, 3-momentum, which is a continuous variable in $\mathbb{R}^3$, plus a set of discrete variables that in our case are going to be particle type (or \textit{flavor}, for both scalar and fermion particles), and helicity (for fermions only). Other discrete quantities like, e.g., electric charge, can be added to the set of discrete variables if needed. In this work, however, we limit ourselves to investigating the quantum nature of flavor and helicity.

We follow closely the formalism described in detail in Refs.\cite{Kowalska:2024kbs,Kowalska:2025qmf}. The Hilbert space for the two-particle system is the tensor product of the momentum~(mom) and discrete-label~(q) spaces:
\be\label{eq:twoHil}
\mathcal{H}_{\textrm{tot}}=\mathcal{H}_{\textrm{mom}}\otimes \mathcal{H}_{\textrm{q}}=L^2(\mathbb{R}^3\otimes \mathbb{R}^3)\otimes\left[\mathbb{C}^{d^2}\otimes\mathbb{C}^4\otimes...\right],
\ee
where $\mathbb{C}^{d^2}=\mathbb{C}^d\otimes \mathbb{C}^d$ and $\mathbb{C}^4=\mathbb{C}^2\otimes \mathbb{C}^2$ are combinations, respectively, of the individual flavor and helicity spaces of the scattering particles. 

The generalized basis vectors in momentum, $| \mathbf{p}_1 \mathbf{p}_2 \rangle$, are normalized as
\be\label{eq:norconv}
\langle\mathbf{p}_1\mathbf{p}_2 | \mathbf{p}_1' \mathbf{p}_2' \rangle
= (2\pi)^6\,4 E_{1} E_{2}\, 
\delta^3(\mathbf{p}_1-\mathbf{p}_1')\, \delta^3(\mathbf{p}_2-\mathbf{p}_2')\,,
\ee
where $E_1$, $E_2$ denote the $0$-th components of momentum 4-vectors $p_{1,2}=(E_{1,2},\mathbf{p}_{1,2})$ in Minkowski space. Given \refeq{eq:norconv} one can introduce
a ``volume'' factor 
\be\label{eq:volume}
V=\langle\mathbf{p}_1\mathbf{p}_2 | \mathbf{p}_1 \mathbf{p}_2 \rangle = 4 E_1 E_2\left[(2\pi)^3\,\delta^3(0)\right]^2,
\ee
which is indeterminate due to the basis vectors 
not being elements of 
$L^2(\mathbb{R}^3\otimes \mathbb{R}^3)$, but rather being tempered distributions acting on its dense subspace.
The basis vectors for the discrete-label space~$\mathcal{H}_{\textrm{q}}$ are normalized trivially as 
\be\label{eq:normfl}
\langle k l,\alpha\beta,... | m n,\gamma\delta,...\rangle=\delta_{km}\,\delta_{ln}\delta_{\alpha\gamma}\,\delta_{\beta\delta}\dots
\ee

For the \textit{separable} initial state before the collision we assume the form
\be\label{eq:genis}
\rho_{\textrm{in}}= |\textrm{in}\rangle\langle\textrm{in}|\otimes
 \tilde{\rho}_{\textrm{in}}\,,
\ee
where $\tilde{\rho}_{\textrm{in}}$ is a $4\times 4$ density matrix. 
In the QED examples discussed in \refsec{sec:QED} the reduced density operator~$\tilde{\rho}_{\textrm{in}}$ is defined in helicity space. Pure state~$|\textrm{in}\rangle$ is instead given by
\be\label{eq:ketin}
|\textrm{in}\rangle=\sum_{l,k=1}^d\int\int\frac{d^3 p_a}{(2\pi)^3\sqrt{2 E_a}}\frac{d^3 p_b}{(2\pi)^3\sqrt{2 E_b}}\,\phi_A^k(\mathbf{p}_a)\phi_B^l(\mathbf{p}_b)|\mathbf{p}_a\mathbf{p}_b;k l\rangle\,,
\ee
where $\phi_{A,B}^{k,l}(\mathbf{p})$ are wave packets in $L^2(\mathbb{R}^3)$ momentum space, sharply peaked about $\mathbf{p}_A,\mathbf{p}_B$ initial-state momenta, 
and $k,l=e,\mu,\tau,...$
is the flavor quantum number. In more general settings $\tilde{\rho}_{\textrm{in}}$
is taken to be any separable $4\times 4$ density operators defined
in $\mathcal{H}_{\textrm{q}}$.

The final state is obtained by applying the $S$-matrix, 
$S=\mathbbm{1}+i T$, to the initial state:
\be\label{eq:gate}
\rho_{\textrm{out}}=S\rho_{\textrm{in}} S^{\dag}\,.
\ee
As is well known, transfer matrix~$T$ can be expressed perturbatively in terms of the scattering amplitude, $\mathcal{M}$. In the basis $\langle \mathbf{p}_i\mathbf{p}_j;m n;\gamma\delta|$, $| \mathbf{p}_a\mathbf{p}_b;k l;\alpha\beta\rangle$ the matrix elements of $T$ read
\be\label{eq:TtoM}
T_{\gamma\delta,\alpha\beta}^{mn,kl}(\mathbf{p}_j\mathbf{p}_j;\mathbf{p}_a\mathbf{p}_b)= (2 \pi)^4
\delta^4(p_a+p_b-p_i-p_j) \mathcal{M}_{\gamma\delta,\alpha\beta}^{mn,kl}(p_a,p_b\to p_i,p_j)\,.
\ee

The scattered state is correctly normalized since $S^{\dag} S=\mathbbm{1}$, which implies the generalized optical theorem,
\be
-i\left(T-T^{\dag} \right) = T^{\dag} T,
\ee
which, at the lowest order in perturbation theory, requires the 
quantities on the l.h.s.~to be computed at one loop. 

In this study we seek to quantify entanglement between the discrete degrees of freedom after the entire momentum space has been traced out. The reduced density matrix in $\mathcal{H}_{\textrm{q}}$ space was derived in Refs.\cite{Kowalska:2024kbs,Kowalska:2025qmf},
\be\label{rho:flav}
\tilde{\rho}_{\textrm{out}}=\tilde{\rho}_{\textrm{in}}+i\Delta\left(\mfw \tilde{\rho}_{\textrm{in}}-\tilde{\rho}_{\textrm{in}}\mfw^{\dag}\right)+\Delta\int d\Pi_2\left(\mathcal{M}\tilde{\rho}_{\textrm{in}} \mathcal{M}^\dag\right),
\ee
where $\mathcal{M}_{\textrm{fw}}\equiv \mathcal{M}(p_A,p_B\to p_A, p_B)$ is the amplitude in the forward direction, we have introduced the common short-hand notation for the phase-space integral,
\be\label{eq:phasespace}
\int d\Pi_2\equiv  \int \int \frac{d^3 p_i}{(2\pi)^3}\frac{1}{2 E_i}\frac{d^3 p_j}{(2\pi)^3}\frac{1}{2 E_j}\,
(2\pi)^4\delta^4(p_A+p_B-p_i-p_j)\,,
\ee
and $\Delta$ 
is a normalization factor of the momentum 
space that parametrizes the 
superposition between the initial- and final-state wave packets. Under the simplifying assumption that the initial momentum state is an element 
$| \mathbf{p}_A \mathbf{p}_B \rangle$ of the generalized basis 
in $L^2(\mathbb{R}^3\otimes \mathbb{R}^3)$, $\Delta$ remains undetermined,
\be\label{eq:Delta}
\Delta=\frac{(2\pi)^4\delta^4(0)}{4 E_A E_B\left[(2\pi)^3\,\delta^3(0)\right]^2}\,.
\ee
In physical situations, however, it
represents a perturbatively small 
but finite quantity\cite{Kowalska:2024kbs,Kowalska:2025qmf}. 

The momentum-reduced density matrix in \refeq{rho:flav} describes a quantum state which is, in general, mixed. Its mixedness is quantified by the trace of the density matrix squared, which reads, in an expansion to the second amplitude order\cite{Kowalska:2025qmf},  
\begin{multline}\label{eq:linen}
\textrm{Tr}\,\tilde{\rho}_{\textrm{out}}^2=
\textrm{Tr}\,\tilde{\rho}_{\textrm{in}}^2-2\Delta\, \textrm{Tr}\int d\Pi_2\left(\tilde{\rho}_{\textrm{in}}\mathcal{M}^{\dag}\mathcal{M}\tilde{\rho}_{\textrm{in}}- \mathcal{M} \tilde{\rho}_{\textrm{in}} \mathcal{M}^{\dag}\tilde{\rho}_{\textrm{in}} \right)\\
+\,\Delta^2\textrm{Tr}\left(2\,\tilde{\rho}_{\textrm{in}}\mfw^{\dag}\mfw \tilde{\rho}_{\textrm{in}}  -\mfw\tilde{\rho}_{\textrm{in}} \mfw\tilde{\rho}_{\textrm{in}} -\mfw^\dagger \tilde{\rho}_{\textrm{in}} \mfw^\dagger\tilde{\rho}_{\textrm{in}}\right).
\end{multline}

In the important case 
of two qubits, an
appropriate measure of entanglement is the concurrence\cite{Hill:1997pfa,Wootters:1997id}. For any $4\times 4$ density matrix~$\rho$ it is defined as
\be\label{eq:conc}
C(\rho)=\max\left[0,\lam_1-\lam_2-\lam_3-\lam_4\right]\,,
\ee
where $\lam_i$ are the square roots of the non-negative eigenvalues (ordered from largest to smallest) of the Hermitian matrix $R=\rho\bar{\rho}$, where $\bar{\rho}$ is the spin-flipped state of $\rho$, defined as
\be
\bar{\rho}=\Sigma\,\rho^\ast\Sigma\,,
\ee
where we define $\Sigma=\sigma_y\otimes\sigma_y$ and $\sigma_y$ indicates the second Pauli matrix. 

It was shown in Ref.\cite{Kowalska:2025qmf} that in the case of a pure initial product state, $\tilde{\rho}_{\textrm{in}}\equiv |A\rangle \langle A|$, the concurrence takes a very simple form at the leading order in perturbation theory,
\be\label{eq:conpr}
C(\tilde{\rho}_{\textrm{out}})=2\,\Delta\,| \bbra\mfw\aket|\simeq 2\,\Delta\, |\bbra\textrm{Re}\,\mfw\aket|\,,
\ee
where $|B\rangle=\Sigma|A^{\ast}\rangle$. Equation~(\ref{eq:conpr}) states that the leading entanglement of a pair of qubits is generated in scattering theory by the forward amplitude. Unitarity implies that, at the leading order, the forward amplitude is real. The contribution to the concurrence of the imaginary part or, equivalently, the scattering cross section, is instead suppressed by one perturbative order. Additionally, it is a straightforward consequence of $\tilde{\rho}_{\textrm{in}}$ and $\tilde{\rho}_{\textrm{out}}$ being Hermitian and $\Delta$ being real that $\textrm{Re}\,\mfw$ appears only in the off-diagonal elements of \refeq{rho:flav}, and it is thus a \textit{coherence} of the density matrix~$\tilde{\rho}_{\textrm{out}}$\cite{Kowalska:2025qmf}.
\paragraph{Unitary quantum channels} 
The formalism outlined above can be straightforwardly rephrased in the language of QIT. The action of the $S$-matrix on the initial-state, cf.~\refeq{eq:gate}, defines a {\it quantum gate}. The evolution of the qubit subsystem, obtained by tracing out the momentum (and all other nonpertinent degrees of freedom)
is referred to as a {\it quantum channel}. Formally, the latter is a map from the initial qubit density matrix to the final one, 
\bea
\mathcal{E}: \mathbb{C}^4 &\to & \mathbb{C}^4\nonumber\\
\tilde{\rho}_{\textrm{in}} & \mapsto & \tilde{\rho}_{\textrm{out}}=\mathcal{E}(\tilde{\rho}_{\textrm{in}}),
\eea
which is linear, trace preserving, and completely positive. 

As was discussed in \refsec{sec:intro}, 
in this work we will be interested in a specific class of quantum channels:
unitary quantum channels,
\be
\mathcal{E}(\tilde{\rho}_{\textrm{in}})=U\tilde{\rho}_{\textrm{in}}U^\dagger,
\ee
where $U$ is a unitary transformation.
The unitarity of a quantum channel has implications on how the scattering event distributes information over the entire system. Given the existence of a unitary channel, if the initial quantum state is separable across the momentum and qubit partitions and the momentum state is pure (which is a reasonable assumption in a scattering event), the final-state density matrix factorizes,
\be\label{eq:fact}
\rho_{\textrm{out}}=\rho_{\textrm{out,mom}}\otimes \tilde{\rho}_{\textrm{out}}\,,
\ee
indicating no correlations, neither quantum nor classical, between qubit and momentum degrees of freedom. Note that \refeq{eq:fact} holds no matter whether the initial qubit state~$\tilde{\rho}_{\textrm{in}}$ is pure or mixed. 

\section{Contact interactions in the 2HDM\label{sec:scalar}}

We begin our investigation of unitary channels with the most straightforward case: contact interactions, which are fully isotropic in phase space. 

Consider the ultrarelativistic scattering of two scalar fields in the 2HDM,
which was recently used as a tool to investigate the potential emergence of global Lagrangian symmetries from the minimization\cite{Carena:2023vjc,Kowalska:2024kbs,Chang:2024wrx,Kowalska:2025qmf} (or maximization\cite{Carena:2025wyh,Chang:2026xif}) of qubit entanglement in the scattering process. While it was shown in Refs.\cite{Kowalska:2024kbs,Chang:2024wrx} that, once all of the scattering processes allowed by the gauge symmetry are considered, the emergence of an additional global symmetry cannot be established through these means, the 2HDM still provides fertile ground for the investigation of QIT properties in perturbative scattering due to its simplicity. 

The 2HDM consists of two scalar doublets of the SU(2)$_L$ 
group of the Standard Model, $H_1$ and $H_2$, which carry  
hypercharge value~$1/2$. Assuming there exists some self-adjoint observable allowing one to distinguish the two doublets after the scattering process, one can construct the qubit out of the doublet's index, $\alpha=1,2$. We dub this index as the field's ``flavor.''  

The full Hilbert space in this case simplifies to
\be\label{eq:red_Hilb}
\mathcal{H}_{\textrm{tot}}=\mathcal{H}_{\textrm{mom}}\otimes\mathcal{H}_{\textrm{q}}=L^2(\mathbb{R}^3\otimes \mathbb{R}^3)\otimes \mathbb{C}^4\,.
\ee

The gauge-invariant scalar potential of the model is given by\cite{Botella:1994cs}
\begin{multline}\label{eq:scapot0}
V(H_1,H_2)=\mu_1^2\,H_1^\dag H_1+\mu_2^2\,H_2^\dag H_2+\left(\mu_3^2\, H_1^\dag H_2+\textrm{H.c.}\right)\\
+\lambda_1\,(H_1^\dag H_1)^2+\lambda_2\,(H_2^\dag H_2)^2+\lambda_3\,(H_1^\dag H_1)(H_2^\dag H_2)+\lambda_4\,(H_1^\dag H_2)(H_2^\dag H_1)\\
+\left(\lambda_5\,(H_1^\dag H_2)^2+\lambda_6\,(H_1^\dag H_1)(H_1^\dag H_2)+\lambda_7\,(H_2^\dag H_2)(H_1^\dag H_2)+\textrm{H.c.}\right),
\end{multline}
where $\mu^2_{1,2,3}\geq 0$ are the squared mass parameters, which we take to be positive, 
and $\lambda_{1,\dots,7}$ denote the dimensionless quartic couplings, which we assume to be real.  The scalar doublets can be explicitly decomposed into charged and neutral components, $H_\alpha=[\,h_\alpha^+,h_\alpha^0\,]^T$.

In the limit of c.o.m.~energy~$E_{\textrm{cm}}$ much larger than the typical scales of the Lagrangian, $E_{\textrm{cm}}\gg \mu_{1,2,3}$, the gauge symmetry of the scalar potential (\ref{eq:scapot0}) allows for five different scattering processes.  The amplitude matrices, $\mathcal{M}^{(n)}_{\alpha\beta,\gamma\delta}$, can be found---expressed at the $n=1$ loop level in the flavor basis $|11\rangle$, $|12\rangle$, $|21\rangle$, $|22\rangle$---in Eqs.~(3.2)-(3.12) of Ref.\cite{Kowalska:2024kbs}.
The amplitude presents 
the same texture in all five combinations of neutral and charged scalar scattering processes enlisted in Ref.\cite{Kowalska:2024kbs}. 

Contact interactions of the neutral and charged components are given by a scattering amplitude that at the tree level is real, symmetric, and independent of the scattering angles. For such an amplitude~$\mathcal{M}^{(0)}$, 
\refeq{eq:linen} reduces to
\be\label{eq:scalpur}
\tr\,\tilde{\rho}_{\textrm{out}}^2=\tr\,\tilde{\rho}_{\textrm{in}}^2-2\Delta\left(\int d\Pi_2-\Delta\right)\tr\left[\tilde{\rho}_{\textrm{in}}(\mathcal{M}^{(0)})^2\tilde{\rho}_{\textrm{in}}-\left(\mathcal{M}^{(0)}\tilde{\rho}_{\textrm{in}}\right)^2\right].
\ee
Equation~(\ref{eq:scalpur}) implies that there exists a limit value of $\Delta$,
\be\label{eq:sep_sc}
\Delta_{\textrm{uni}}=\int d\Pi_2\,,
\ee
that is trace-squared conserving. 
As was discussed in Ref.\cite{Kowalska:2025qmf}, the second set of addends on the r.h.s.~of \refeq{eq:linen} is proportional to the scattering cross section and contributes positively to the linearized entropy. Conversely, the third set of addends 
on the r.h.s.~of \refeq{eq:linen}
quantifies the counterweight effect of
the coherences of the final-state density matrix. A physical interpretation of \refeq{eq:sep_sc} thus possibly rests on $\Delta=\Delta_{\textrm{uni}}$ being the limit value of the initial-state angular spread maximizing the effects of the coherences of the two-qubit system against the environment~(momentum) in the scattering event.

As was mentioned in \refsec{sec:forma}, $\Delta$ parametrizes the degree of overlap between the initial- and final-state wave packets. As such, \refeq{eq:sep_sc} implies that the wave packets of the $|\textrm{in}\rangle$ state have been somehow prepared to be fully isotropic so to mimic the probability distribution of the final state. Such a situation is obviously not physical in a collider event. It can be taken, however, as a mathematical limit, under which the density matrix can be written as
\be
\tilde{\rho}_{\textrm{out}}=U \tilde{\rho}_{\textrm{in}} U^\dagger\,,
\ee
where
\be\label{eq:scalun}
U=e^{i\, \Delta_{\textrm{uni}} \mathcal{M}^{(0)}}\simeq\mathbbm{1}+i\, \Delta_{\textrm{uni}} \mathcal{M}^{(0)}-\frac{1}{2}(\Delta_{\textrm{uni}}\mathcal{M}^{(0)})^2+\dots\,,
\ee
up to the second order in the perturbative expansion. In other words, there exist a limit solution under which the scattering process defines a unitary quantum channel in $\mathbb{C}^4$.

In order to quantify the amount of entanglement generated in the flavor system by the quantum channel defined in \refeq{eq:scalun}, one needs the parametric dependence of the amplitude. Let us focus for concreteness on one particular process, $h^+h^0\to h^+h^0$, keeping in mind that the discussion for the other four processes follows the same lines. For simplicity, we will also assume the $\mathbbm{Z}_2$-symmetric version of the 2HDM, in which the couplings $\lambda_6$ and $\lambda_7$ are set to zero, and $\lam_2=\lam_1$. We recall the tree-level scattering matrix in the flavor basis \{$|11\rangle$, $|12\rangle$, $|21\rangle$, $|22\rangle$\}\cite{Kowalska:2024kbs},
\be\label{eq:scat0p}
i\mathcal{M}^{(0)}(h^+h^0\to h^+h^0)=-i\left(\begin{array}{cccc} 
2\lambda_1 & 0 & 0 & 2\lambda_5  \\ 
0 & \lambda_3 & \lambda_4 & 0 \\
0 & \lambda_4 & \lambda_3 & 0 \\
2\lambda_5 & 0 & 0 & 2\lambda_1
\end{array}\right).
\ee

We will analyze three families of separable mixed initial states that depend on a free parameter $0\leq b\leq 1$. Let us first consider a generic diagonal state, $\tilde{\rho}^{(\pm\pm)}_{\textrm{in},l}$, such that each of the four possible combinations of superscripts correspond to a computational basis vector and such that one of the combinations reads
\be\label{eq:long_sca}
\tilde{\rho}^{(++)}_{\textrm{in},l}(b)=\frac{1}{4}\left(\begin{array}{cccc}
\left(1+b \right)^2  & 0 & 0 & 0 \\
0 & 1-b^2 & 0 & 0 \\
0 & 0 & 1-b^2 & 0 \\
0 & 0 & 0 & \left(1-b \right)^2
\end{array}\right)\,.
\ee
We take \refeq{eq:long_sca} to be our first benchmark family and we take the second to be $\tilde{\rho}^{(+-)}_{\textrm{in},l}(b)$, which is derived from 
$\tilde{\rho}^{(++)}_{\textrm{in},l}(b)$ after exchanging elements
$11\leftrightarrow 22$ and $33\leftrightarrow 44$.
There exist two additional states that can be constructed out of $\tilde{\rho}^{(++)}_{\textrm{in},l}(b)$: by switching $11\leftrightarrow 33$, $22\leftrightarrow 44$; and by switching $11\leftrightarrow 44$, $22\leftrightarrow 33$. 
As we shall see in \refsec{sec:QED}, these states show identical texture to the four helicity states of QED with longitudinal polarization.

It is straightforward to see that the two selected initial states are pure only when $b=1$, where they become $\tilde{\rho}^{(++)}_{\textrm{in},l}=|11\rangle \langle 11|$ and $\tilde{\rho}^{(+-)}_{\textrm{in},l}=|12\rangle \langle 12|$. In general, though, they are mixed,  
\be\label{eq:tr2}
\textrm{Tr}\,(\tilde{\rho}^{(++)}_{\textrm{in},l})^2=\textrm{Tr}\,(\tilde{\rho}^{(+-)}_{\textrm{in},l})^2=\frac{1}{4}\left(1+b^2 \right)^2,
\ee 
and they are maximally mixed at $b=0$. 

The third family of mixed initial states we choose as our benchmark features nonzero coherences:
\be\label{eq:tran_sca}
\tilde{\rho}_{\textrm{in},t}(b)=\frac{1}{4}\left(\begin{array}{cccc}
1 & b & b & b^2 \\
b  & 1 &  b^2 & b \\
b & b^2 & 1 & b \\
b^2  & b  & b  & 1
\end{array}\right),
\ee
which bears resemblance, as we shall see in \refsec{sec:QED}, to one of the four QED states with transverse polarization. The states in Eqs.~(\ref{eq:long_sca}) and~(\ref{eq:tran_sca}) feature identical spectrum and mixedness, i.e., they are pure at $b=1$ and maximally mixed at $b=0$.

After applying the $S$-matrix, imposing optical theorem, tracing out the momentum degrees of freedom, and expanding to order~$\mathcal{O}(\lam^2)$, 
the reduced density matrix takes in all cases a simple form:
\be\label{eq:sc_long}
\tilde{\rho}_{\textrm{out},l(t)}=\tilde{\rho}_{\textrm{in},l(t)}+b\, \Delta\,\delta\tilde{\rho}_{l(t)}\,,
\ee
where 
\be\label{eq:outdm1}
\delta\tilde{\rho}_l^{(++)}=-2 \lam_5\left(\begin{array}{cccc}
0 & 0 & 0 & -i \\
0 & 0 & 0 & 0 \\
0 & 0 & 0 & 0 \\
i & 0 & 0 & 0
\end{array}\right)-\frac{(2\lam_5)^2}{8 \pi}\left(\begin{array}{cccc}
1 & 0 & 0 & 0 \\
0 & 0 & 0 & 0 \\
0 & 0 & 0 & 0 \\
0 & 0 & 0 & -1
\end{array}\right)+\mathcal{O}(\lam^3),
\ee
\be\label{eq:outdm2}
\delta\tilde{\rho}_l^{(+-)}=-\lam_4\left(\begin{array}{cccc}
0 & 0 & 0 & 0 \\
0 & 0 & -i & 0 \\
0 & i & 0 & 0 \\
0 & 0 & 0 & 0
\end{array}\right)-\frac{\lam_4^2}{8 \pi}\left(\begin{array}{cccc}
0 & 0 & 0 & 0 \\
0 & 1 & 0 & 0 \\
0 & 0 & -1 & 0 \\
0 & 0 & 0 & 0
\end{array}\right)+\mathcal{O}(\lam^3),
\ee
\be\label{eq:outtm}
\delta\tilde{\rho}_t=-c_\lam\left(\begin{array}{cccc}
0 & i & i & 0 \\
-i & 0 & 0 & -i \\
-i & 0 & 0 & -i \\
0 & i & i & 0
\end{array}\right)-\frac{c_\lam^2}{4\pi}\left(\begin{array}{cccc}
0 & 1 & 1 & 0 \\
1 & 0 & 0 & 1 \\
1 & 0 & 0 & 1 \\
0 & 1 & 1 & 0
\end{array}\right)+\mathcal{O}(\lam^3),
\ee
and we have defined 
\be\label{eq:clam}
c_{\lam}\equiv \frac{1}{4}\left(2\lam_1+2\lam_5-\lam_3-\lam_4\right).
\ee
Note that maximally mixed states ($b=0$) remain unchanged after the scattering process, while for the others the purity decreases, 
\be\label{eq:pur_sc}
\tr\,\tilde{\rho}_{\textrm{out},l(t)}^2=\tr\,\tilde{\rho}_{\textrm{in},l(t)}^2-2\, b^2 \alpha^2_{l(t)}\Delta\left(\frac{1}{8\pi}-\Delta \right)+\mathcal{O}(\alpha^4)
\ee
(recall that $\Delta$ is typically much smaller than the full solid angle opening). In \refeq{eq:pur_sc} we define 
\bea
\alpha_l^{(++)}&=&2\lam_5\,,\\
\alpha_l^{(+-)}&=&\lam_4\,,\\
\alpha_t&=& 2\,c_{\lam}\,.
\eea

Given the scattered states~(\ref{eq:sc_long}),
the concurrence can be calculated at the leading order,
\be\label{eq:conc_sc}
C(b)=\max\left[0,2\,b \left|\alpha_{l(t)}\right| \Delta-\frac{1}{2}\left(1-b^2\right)\right].
\ee
Maximally mixed states ($b=0$) lead to $C(0)=0$, 
independently of the coupling strength, which means that no entanglement between the flavor qubits is ever generated at the leading order in $\alpha$. Pure states ($b=1$) lead instead to the highest amount of entanglement that the $S$-matrix can generate, $C(1)=2|\alpha_{l(t)}|\Delta+\mathcal{O}(\alpha^2)$.

As we have discussed above, $\Delta=\Delta_{\textrm{uni}}$ defines a unitary channel over $\mathbb{C}^4$. Equation~(\ref{eq:sep_sc}) can be integrated straightforwardly in the ultrarelativistic regime and yields 
$\Delta_{\textrm{uni}}=1/(8\pi)$, in agreement with
\refeq{eq:pur_sc}. One can verify explicitly by expanding the eigenvalues to $\mathcal{O}(\lam^2)$ that the spectrum of the qubit density matrix is conserved at the desired order,
\be
\textrm{spec}\left[\tilde{\rho}_{\textrm{out},l(t)}\right]=\textrm{spec}\left[\tilde{\rho}_{\textrm{in},l(t)}\right]\quad \textrm{at }\Delta=\Delta_{\textrm{uni}}\,.
\ee

The unitarity transformation~(\ref{eq:scalun}) allows us to  explore some specific QIT properties of the two-qubit density matrix. In particular, it was shown in Ref.\cite{Verstraete:2001okv} that 
the maximum concurrence that can be achieved by applying a unitary transformation to a $\mathbb{C}^4$ mixed state is given by
\be
C_{\textrm{max}}=\max\left[0,\xi_1-\xi_3-2\sqrt{\xi_2\,\xi_4}\right],
\ee
where $\xi_i$ are the eigenvalues of the initial-state density matrix in decreasing order. Given that all three initial-state benchmarks have the same spectrum, 
one finds a common value,
\be\label{eq:maxent}
C_{\textrm{max}}=\max\left[0,\frac{1}{2}\left(b\left(1+b\right)-\left(1-b\right)\sqrt{1-b^2} \right)\right].
\ee
Note that \refeq{eq:maxent} implies the existence of a lower bound for the purity of 
the initial state, below which the mixed state cannot be entangled by a unitary transformation. This property of a quantum state is sometimes dubbed in the literature as {\it absolute separability}\cite{Verstraete:2001okv}. Given our benchmark states, one finds $b_{\textrm{abs}}=0.397$. 

At any $b>b_{\textrm{abs}}$, the amount of qubit-qubit entanglement generated by the scattering event is determined by the interaction strength. Plugging  $\Delta_{\textrm{uni}}=1/(8\pi)$ into \refeq{eq:conc_sc}, 
one can extract an upper bound on the coupling constants by applying the physicality condition $C\leq C_{\textrm{max}}$:
\be\label{eq:sc_uni_bound1}
 |\alpha_{l(t)}| \leq 
\frac{2\pi}{b}\left(1+b-(1-b)\sqrt{1-b^2}\right).
\ee
This limit provides the maximal value of the coupling(s) that guarantees the internal consistency of the concurrence at a given perturbative order. Therefore, we are going to refer to it as a {\it perturbative unitarity} bound. Conversely, the entanglement generating condition $C > 0$ yields the minimal interaction strength necessary to overcome the mixedness of the initial state: 
\be\label{eq:sc_uni_bound2}
|\alpha_{l(t)}| > 2\pi\frac{\left(1-b^2\right)}{b}\,.
\ee

For a pure initial state ($b=1$), \refeq{eq:sc_uni_bound1} yields $|\alpha_{l(t)}| \leq 4\pi$.
Applying it to all the initial states considered in this section, one obtains:
\be\label{eq:unibou}
|\lam_4|\leq 4\pi\,,\qquad |\lam_5|\leq 2\pi\,,\qquad |2\lam_1+2\lam_5-\lam_3-\lam_4|\leq 8\pi\,.   
\ee

 \begin{figure}[t]
	\centering%
		\includegraphics[width=0.65\textwidth]{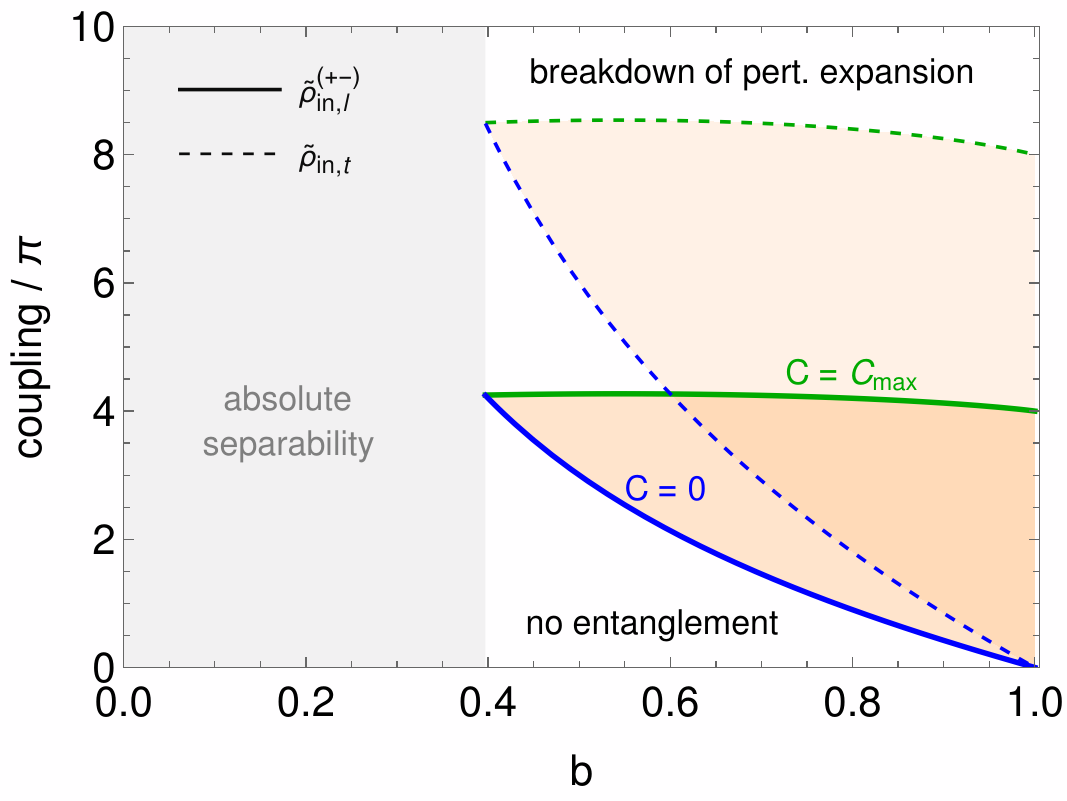}
\caption{Bounds on the effective scalar coupling set by the maximal (green) and minimal (blue) value of the concurrence~(\ref{eq:conc_sc}) at $\Delta=\Delta_{\textrm{uni}}$. Dashed lines apply to the coupling combination ${|2\lam_1+2\lam_5-\lam_3-\lam_4|}$, solid lines apply to $\lam_4$ only. Orange shading indicates the parameter space where nonzero entanglement between the flavor qubits can be generated.}
\label{fig:scalun}
\end{figure}

In \reffig{fig:scalun} we present upper bounds on the coupling/coupling product $\alpha$ stemming from the properties of the concurrence at $\Delta=\Delta_{\textrm{uni}}$, as a function of the initial state purity $b$. In green (solid \textit{and} dashed) we show the values of the couplings for which $C(b)$ reaches the maximal value that can be obtained from the initial state $\tilde{\rho}_{\textrm{in},l(t)}(b)$. Solid line corresponds to the diagonal initial state $\tilde{\rho}^{(+-)}_{\textrm{in},l}(b)$, 
while dashed line corresponds to the initial state $\tilde{\rho}_{\textrm{in},t}$ in \refeq{eq:tran_sca}.
Above the green lines the concurrence becomes unphysical, signaling the breakdown of the perturbative expansion. 

The blue lines (solid and dashed) in \reffig{fig:scalun} correspond to the condition $C(b)=0$. In the parameter space below the blue lines no entanglement between the flavor qubits can be generated, as the interaction is too weak to overcome the mixedness of the initial state. As the parameter~$b$ decreases, larger and larger couplings are required to induce quantum correlations in the flavor system. Eventually, at $b\approx 0.4$ the 
absolute separability bound is reached, indicated in \reffig{fig:scalun} as a shaded gray region. Finally, the orange-shaded regions between the green and blue lines restrict the parameter space where non-trivial correlations between flavor degrees of freedom are generated by the scattering amplitude. 

To close this section, let us compare the bounds (\ref{eq:unibou}) with partial-wave unitarity bounds, which result from the requirement that the outgoing probability flux in a scattering process does not exceed the incoming flux. They are commonly phrased in terms of constraints on the partial-wave amplitudes~$a_l$, defined as  
\be
a_l(s)=\frac{1}{32\pi}\int_{-1}^1d(\cos\theta)\, P_l(\cos\theta)\mathcal{M}(\theta,s)\,,
\ee
where $\mathcal{M}$ is a scattering amplitude, $P_l$ are Legendre polynomials, and the high-energy limit is assumed. For a purely real tree-level amplitude the bounds read
\be\label{eq:pwave}
|\textrm{Re}(a_l)|\leq \frac{1}{2}\,.
\ee
Note that if $\mathcal{M}$ is a matrix, inequality~(\ref{eq:pwave}) applies to the eigenvalues of $a_l$, the largest eigenvalue giving the strongest constraint. For the amplitude defined in \refeq{eq:scat0p}, the eigenvalues of the $s$-wave partial wave~$a_0$ are given by
\be
\textrm{Eigenvalues}(a_0)=\left\{
\frac{1}{16\pi}\left(\lam_3\pm\lam_4\right),\,
\frac{1}{8\pi}\left(\lam_1\pm\lam_5\right)\right\}\,.
\ee
Assuming that all the couplings are positive, the partial-wave unitarity bounds read
\be\label{eq:pwbound_sca}
|\lam_3+\lam_4|\leq 8\pi\,,\qquad |\lam_1+\lam_5|\leq 4\pi\,.
\ee

We point out that the bounds in \refeq{eq:pwbound_sca}
agree individually with the bounds~(\ref{eq:unibou}) extracted from the concurrence. This is encouraging and validates the consistency of our method, which, we recall, is based exclusively on the QIT properties of the two-qubit system after scattering.

\section{Quantum electrodynamics\label{sec:QED}}

We now consider cases where the amplitude depends explicitly on the momentum coordinates. This is the familiar situation, for example, in QED. Quantum spin correlations in QED were analyzed in a number of papers\cite{Cervera-Lierta:2017tdt,Fonseca:2021uhd,Blasone:2024dud,Blasone:2024jzv,Blasone:2025tor,Blasone:2025ddi,Martin:2025hzm,Guo:2026yhz, Fang:2026ddi,Fonseca:2026nma}. Our goal is to quantify helicity entanglement generated in the unitary channel, following the steps performed in \refsec{sec:scalar} for contact interactions.

We adopt in this work the Jacob-Wick convention\cite{Jacob:1959at}, based on which one can define a helicity basis in $\mathbb{C}^4$ that remains independent of the momentum degrees of freedom in both the initial and final state. We work in the computational basis of the two-qubit system: $|\alpha\beta\rangle =|RR\rangle,|RL\rangle,|LR\rangle,|LL\rangle$, with $R(L)$ indicating 
that the 
spin of the particle or antiparticle is aligned(antialigned) with its direction of motion. 

We construct the scattering amplitudes in the c.o.m.~frame and take the electron mass to be negligible with respect to $E_{\textrm{cm}}$ and all other masses in the process.
The $T$ matrix defined in \refeq{eq:TtoM} takes the form
\be\label{eq:T-matr}
T_{\gamma\delta,\alpha\beta}^{mn,kl}(\mathbf{p}_i\mathbf{p}_j;\mathbf{p}_A\mathbf{p}_B)
=(2\pi)^4 \delta(E-E_{\textrm{cm}})\, \delta^3(\mathbf{P})\, \mathcal{M}_{\gamma\delta,\alpha\beta}^{mn,kl}(E,\mathbf{P},\theta,\phi)\,,
\ee
where in \refeq{eq:T-matr} we have defined the usual c.o.m.~coordinates,
\be
\mathbf{P}=\mathbf{p}_i+\mathbf{p}_j\,,\quad |\mathbf{p}_{i,j}|=\sqrt{E^2-m_{m,n}^2}\,,
\ee
$m_{m,n}$ are the masses of the final-state flavors, $\theta$ specifies the polar angle of relative momentum $\mathbf{p}=(m_m+m_n)^{-1}(m_n \mathbf{p}_i-m_{m}\mathbf{p}_j)$ with respect to the $z$-direction, and $\phi$ specifies the azimuthal angle of $\mathbf{p}$ with respect to the $x$-direction.

\subsection{Moeller scattering\label{sec:Moeller}}

Let us first consider QED with one flavor. By collapsing the flavor quantum number the Hilbert space becomes isomorphic to \refeq{eq:red_Hilb}.

We focus in this section on Moeller scattering, $e^- e^-\to e^- e^-$. At the tree level, the amplitude includes the $t$ and $u$ channels. We define initial 4-momenta as $p_a(e^-)$, $p_b(e^-)$, and the final momenta as $p_i(e^-)$, $p_j(e^-)$. The amplitude is well known, 
\begin{multline}
i\mathcal{M}^{(0)}=\frac{i e^2}{\left(p_a-p_i \right)^2}\left[\bar{u}(p_i)\gamma^{\mu} u(p_a) \right] \left[\bar{u}(p_j)\gamma_{\mu} u(p_b) \right]\\
-\frac{i e^2}{\left(p_a-p_j \right)^2}\left[\bar{u}(p_j)\gamma^{\mu} u(p_a) \right] \left[\bar{u}(p_i)\gamma_{\mu} u(p_b) \right]\,.
\end{multline}

Neglecting the electron mass, inserting the four helicity projectors, and working in c.o.m.~coordinates we recast the amplitude in the $\{|RR\rangle$, $|RL\rangle$, $|LR\rangle$, $|LL\rangle\}$ basis, obtaining
\be\label{eq:Moe_amp}
\mathcal{M}^{ee,ee}_{\textrm{(tree)}}=2 e^2\left(\begin{array}{cccc} 
\frac{4}{c_{\theta}^2-1} & 0 & 0 & 0\\ 
0 & \frac{1+c_{\theta}}{c_{\theta}-1} & \frac{\left(1-c_{\theta}\right)e^{-2 i \phi}}{1+c_{\theta}} & 0 \\
0 &  \frac{\left(1-c_{\theta}\right)e^{2 i \phi}}{1+c_{\theta}} & \frac{1+c_{\theta}}{c_{\theta}-1} & 0 \\
0 & 0 & 0 & \frac{4}{c_{\theta}^2-1}
\end{array}\right),
\ee
where we have defined $c_{\theta}\equiv \cos\theta$ for economy of notation. Note that the amplitude features the well-known singular behavior at $c_{\theta}=\pm 1$. 
In this section, we will consider transversely polarized incoming states with polarizations along the $x$-axis, so that incoming $e^- e^-$ momenta are along the $z$-axis, their helicities are along the $x$-axis, and the outgoing particles can in full generality be outside of this $x-z$ plane, 
with their azimuthal angle given by $\phi$.

As was discussed in Refs.\cite{Kowalska:2024kbs,Kowalska:2025qmf}, 
working with the tree-level amplitude does not suffice to guarantee that the $S$-matrix is unitary, even at the leading order. As is well known, one ought to  take into account the imaginary part of the forward scattering amplitude, which can be generated at the 1-loop level and equals the integral over phase space of the absolute value squared of the tree-level amplitude. Explicitly, given \refeq{eq:Moe_amp}, one can write,
\begin{multline}\label{eq:Moeller_loop}
2\, \textrm{Im}\,\mathcal{M}_{\textrm{fw}}^{(\textrm{1-loop})}=\frac{1}{32 \pi^2}\int_{0}^{1-\epsilon}d(\cos\theta)\int_0^{2\pi}d\phi\, |\mathcal{M}|^2 \\
\simeq e^4\left(\begin{array}{cccc} 
-\frac{1}{2\pi}+\frac{1}{\pi \epsilon}-\frac{1}{\pi}\ln\frac{\epsilon}{2} & 0 & 0 & 0\\ 
0 & \frac{1}{\pi \epsilon}+\frac{1}{\pi}\ln\frac{\epsilon}{2}  & 0 & 0 \\
0 & 0 & \frac{1}{\pi \epsilon}+\frac{1}{\pi}\ln\frac{\epsilon}{2}  & 0 \\
0 & 0 & 0 & -\frac{1}{2\pi}+\frac{1}{\pi \epsilon}-\frac{1}{\pi}\ln\frac{\epsilon}{2}
\end{array}\right),
\end{multline}
where the phase space in \refeq{eq:Moeller_loop} is symmetrized to account for the identical particles in the final state, and we have introduced an angular cutoff~$\epsilon$ to regularize the singular behavior of the forward amplitude. As we shall see below, the cutoff dependence is expected to cancel out, order by order, from the final-state density matrix.  

We focus here on initial states with \textrm{transverse} ($x$-direction) polarization, which allow for a comprehensive treatment of initial states with nonzero coherences in the density matrix. 
Following, for example, the discussion in Ref.\cite{Fang:2026ddi},
we construct four possible helicity space density matrices,
\be\label{eq:tra_in}
\tilde{\rho}_{\textrm{in}}^{(\pm \pm)}=\frac{1}{2}\left(\mathbbm{1}_2\pm b_x \sigma_x \right)\otimes \frac{1}{2}\left(\mathbbm{1}_2\pm b_x \sigma_x \right),
\ee
where $\sigma_x$ is the usual Pauli matrix and $0\leq b_x \leq 1$ parametrizes the degree of polarization. In order to make contact with the state defined in \refeq{eq:tran_sca}, we select, without loss of generality, the corresponding possibility in \refeq{eq:tra_in}, 
\be\label{eq:tra_in1}
\tilde{\rho}_{\textrm{in}}^{(++)}=\frac{1}{4}\left(\begin{array}{cccc}
1 & b_x & b_x & b_x^2 \\
b_x  & 1 &  b_x^2 & b_x  \\
b_x  & b_x^2 & 1 & b_x \\
b_x^2  & b_x  & b_x  & 1
\end{array}\right).
\ee
As in the scalar case, the initial state is typically mixed,  
\be
\textrm{Tr}\,(\tilde{\rho}_{\textrm{in}}^{(++)})^2=\frac{1}{4}\left(1+b_x^2 \right)^2\leq 1\,,
\ee
cf.~the equivalent expression in \refsec{sec:scalar}.

We apply the $S$-matrix to the initial state, impose optical theorem, trace out the momentum degrees of freedom, and expand the result to $\mathcal{O}(e^4)$. In \refeq{rho:flav}, the phase-space integrals of 
the matrix elements characterized by a symmetric distribution in $\cos\theta$ are symmetrized to avoid the double counting of identical particles. One finds
\be\label{eq:Mo_trans}
\tilde{\rho}_{\textrm{out}}^{(++)}=\tilde{\rho}_{\textrm{in}}^{(++)}+b_x\, \Delta\,\delta\tilde{\rho}^{(++)},
\ee
where 
\be\label{eq:corr}
\delta\tilde{\rho}^{(++)}=-e^2 \left(\begin{array}{cccc}
0 & i & i & 0 \\
-i & 0 & 0 & -i \\
-i & 0 & 0 & -i \\
0 & i & i & 0
\end{array}\right)-\frac{e^4}{16\pi}\left(\begin{array}{cccc}
0 & 1 & 1 & 0 \\
1 & 0 & 0 & 1 \\
1 & 0 & 0 & 1 \\
0 & 1 & 1 & 0
\end{array}\right)+\mathcal{O}(e^6).
\ee
Equation~(\ref{eq:corr}) is completely analogous to \refeq{eq:outtm}.
As was the case in \refsec{sec:scalar}, it yields a well defined density matrix up to the indicated order in perturbation theory. Note, importantly, that the cutoff~$\epsilon$, introduced in \refeq{eq:Moeller_loop}, has disappeared. The density matrix is effectively ``renormalized'' at the leading order through the application of the optical theorem.

The final-state purity can be computed from 
\refeq{eq:linen} and it
shows an expression similar to \refeq{eq:pur_sc},
\be\label{eq:Mo_tr2}
\textrm{Tr}\,(\tilde{\rho}_{\textrm{out}}^{(++)})^2=\textrm{Tr}\,(\tilde{\rho}^{(++)}_{\textrm{in}})^2
-2\, b_x^2 e^4 \Delta \left(\frac{1}{8\,\pi}-4\,\Delta\right)+\mathcal{O}(e^8)\,,
\ee
where, as was explained in \refsec{sec:scalar}, the second addend on the r.h.s.~quantifies the positive contribution of the scattering cross section to the entropy while the third addend shows the counteraction of the final-state coherences\cite{Kowalska:2025qmf}. 
By imposing conservation of the trace in \refeq{eq:Mo_tr2} one finds $\Delta_{\textrm{uni}}=1/(32\pi)$. 

The concurrence takes  at $O(e^2)$ a form analogous to \refeq{eq:conc_sc},
\be\label{eq:conc_mo}
C(b_x)=\max\left[0,4\,b_x e^2 \Delta-\frac{1}{2}\left(1-b_x^2\right)\right].
\ee

Following the discussion in \refsec{sec:scalar}, we set $\Delta=\Delta_{\textrm{uni}}$ and attempt to establish a unitary channel in helicity space. The unitary transformation 
\be
U=e^{i \Delta_{\textrm{uni}} K}\simeq \mathbbm{1}+i \Delta_{\textrm{uni}} K-\frac{1}{2}\left( \Delta_{\textrm{uni}} K \right)^2+\dots\,,
\ee
where the matrix $K$ is defined as
\be\label{eq:kmatrix}
K=e^2\begin{pmatrix}
    -1 & 0 & 0 & -1 \\
    0 & 1 & 1 & 0 \\
    0 & 1 & 1 & 0 \\
        -1 & 0 & 0 & -1 \\
\end{pmatrix},
\ee
conserves the spectrum for the parametric states $\tilde{\rho}_{\textrm{in}}^{(++)}$. 
One can thus extract a perturbative unitarity bound on the coupling constant equivalent to \refeq{eq:sc_uni_bound1},
\be\label{eq:Mo_uni_bound}
 e^2 \leq 
\frac{4\,\pi}{b_x}\left(1+b_x-\sqrt{\left(1-b_x\right)^3\left(1+b_x\right)}\right).
\ee
For a pure state it reads $e^2 \leq 8\pi$.

The bound in \refeq{eq:Mo_uni_bound} is finite at the leading order in perturbation theory. One can thus see that, under conditions that allow us to interpret the scattering process as a unitary transformation in qubit-qubit space, we are able to overcome some of the hurdles undermining the traditional partial-wave expansion approach, which, 
in the case of Moeller scattering, typically requires a renormalization scheme to absorb the singularities of forward scattering in the $t$- and $u$-channel\cite{FuentesZamoro:2025exp,Plestid:2026pzn,Lippstreu:2026xfw}. 

We also point out, on the other hand, that there is no \textit{a priori} guarantee that a quantum channel could be readily established over the full $\mathbb{C}^4$ Hilbert space. To this regard, let us briefly analyze the case of Moeller scattering from an initial state with longitudinal ($z$-direction) polarization. Consider an initial state partially polarized in the $z$-direction, with density matrix as discussed
below \refeq{eq:long_sca},
\be\label{eq:long_in}
\tilde{\rho}_{\textrm{in}}^{(+-)}=\frac{1}{2}\left(\mathbbm{1}_2+ b_z \sigma_z \right)\otimes \frac{1}{2}\left(\mathbbm{1}_2- b_z \sigma_z \right),
\ee
where $\sigma_z$ is the usual Pauli matrix. The final state will assume an expression analogous to \refeq{eq:outdm2},
with the coupling replaced by the corresponding entry in \refeq{eq:Moe_amp}. To extract that, we recall a direct consequence of \refeq{rho:flav}: the coherences of the density matrix in \refeq{eq:outdm2} are generated at $\mathcal{O}(\lam)$ from the tree-level forward amplitude. Thus, given amplitude~(\ref{eq:Moe_amp}), it follows that the corresponding coherences of the final-state density matrix are zero at this order:
\be
2 e^2 \frac{1-c_{\theta}}{1+c_{\theta}}\big|_{c_{\theta}=1}=0\,.
\ee
As a straightforward consequence of  
\refeq{eq:linen} one can see that if $\mathcal{M}_{\textrm{fw}}=0$
there is no nontrivial trace conserving solution $\Delta=\Delta_{\textrm{uni}}$, so that a unitarity 
transformation in $\mathbb{C}^4$ cannot be established for initial states with longitudinal polarization.

In other words,  for a longitudinally polarized helicity state, a quantum channel in the helicity space cannot be unitary. Consequently, momentum and helicity Hilbert spaces do not factorize and nontrivial quantum correlations between momentum and helicity degrees of freedom are always present. We thus need to conclude that Moeller scattering, unlike contact scalar interactions discussed in \refsec{sec:scalar}, only allows a unitary channel with respect to the family of states with transverse polarization, at least at the considered perturbative order. Interestingly, it was shown in Ref.\cite{Kowalska:2025qmf} that the absence of coherences in the final-state density matrix also implies that
the concurrence is zero at $\mathcal{O}(e^2)$, so that a perturbativity bound on the coupling constant cannot be extracted at this order starting from a state with longitudinal polarization. 

\subsection{Bhabha scattering\label{sec:Bhabha}}

Let us now consider Bhabha scattering, $e^- e^+ \to e^- e^+$. At the tree level, it proceeds through the $s$-channel and the $t$-channel. 
We define the initial momenta as $p_a(e^-)$, $p_b(e^+)$, and the final momenta as $p_i(e^-)$, $p_j(e^+)$. The tree-level amplitude reads
\begin{multline}
i\mathcal{M}^{(0)}=\frac{i e^2}{\left(p_a+p_b \right)^2}\left[\bar{v}(p_b)\gamma^{\mu} u(p_a) \right] \left[\bar{u}(p_i)\gamma_{\mu} v(p_j) \right]\\
-\frac{i e^2}{\left(p_a-p_i \right)^2}\left[\bar{u}(p_i)\gamma^{\mu} u(p_a) \right] \left[\bar{v}(p_b)\gamma_{\mu} v(p_j) \right]\,.
\end{multline}

Neglecting the electron mass, inserting the four helicity projectors, and working in c.o.m.~coordinates we recast the amplitude in the $\{|RR\rangle$, $|RL\rangle$, $|LR\rangle$, $|LL\rangle\}$ basis, obtaining
\be\label{eq:Bhabha_tree}
\mathcal{M}^{ee,ee}_{(\textrm{tree})}=e^2\left(\begin{array}{cccc} 
\frac{4}{1-c_{\theta}} & 0 & 0 & 0\\ 
0 & \frac{\left(1+c_{\theta}\right)^2}{1-c_{\theta}} & -\left(1-c_{\theta}\right)e^{-2 i \phi} & 0 \\
0 &  -\left(1-c_{\theta}\right)e^{2 i \phi} & \frac{\left(1+c_{\theta}\right)^2}{1-c_{\theta}} & 0 \\
0 & 0 & 0 & \frac{4}{1-c_{\theta}}
\end{array}\right)\,,
\ee
where, again, $c_{\theta}\equiv \cos\theta$. We recall that the $t$-channel presents a forward singularity at $c_{\theta}=1$. 

We focus once more on initial states with transverse polarization since, in analogy to Moeller scattering in \refsec{sec:Moeller}, states with longitudinal polarization do not give rise to a nonzero concurrence at the leading order, $\mathcal{O}(e^2)$. 
The initial state density matrix is thus taken to be \refeq{eq:tra_in1}.

We apply the $S$-matrix, impose optical theorem, trace out the momentum degrees of freedom, and expand to $\mathcal{O}(e^4)$. Interestingly, the reduced density matrix shows a richer texture than in the cases presented in \refsec{sec:scalar} and \refsec{sec:Moeller}. One finds,
\be\label{eq:Bb_trans}
\tilde{\rho}_{\textrm{out}}^{(++)}=\tilde{\rho}_{\textrm{in}}^{(++)}+b_x\,\Delta\,\delta\tilde{\rho}^{(++)}\,,
\ee
where, in this case,
\be\label{eq:corr_Bb}
\delta\tilde{\rho}^{(++)}=-e^2 \left(\begin{array}{cccc}
0 & -i & -i & 0 \\
i & 0 & 0 & i \\
i & 0 & 0 & i \\
0 & -i & -i & 0
\end{array}\right)-\frac{e^4}{6\pi}\left(\begin{array}{cccc}
0 & 1 & 1 & 0 \\
1 & 0 & 0 & 1 \\
1 & 0 & 0 & 1 \\
0 & 1 & 1 & 0
\end{array}\right)
-\frac{e^4}{6\pi}\left(\begin{array}{cccc}
0 & 0 & 0 & 0 \\
0 & 0 & \frac{b_x}{4} & 0 \\
0 & \frac{b_x}{4} & 0 & 0 \\
0 & 0 & 0 & 0
\end{array}\right)\dots
\ee
Note that, differently to Moeller scattering, the interaction has generated additional coherences of the density matrix at $\mathcal{O}(e^4)$, proportional to the degree of initial state polarization $b_x$. 
As these coherences appear at higher order, they will not affect 
the concurrence, which is determined at lowest order by the first addend on the r.h.s.~of \refeq{eq:corr_Bb} and thus remains unaltered with respect to \refeq{eq:conc_mo}. Rather, the additional coherences increase the mixedness of the final state, which is now given by
\be\label{eq:Bb_tr2}
\textrm{Tr}\,(\tilde{\rho}_{\textrm{out}}^{(++)})^2=\textrm{Tr}\,(\tilde{\rho}^{(++)}_{\textrm{in}})^2
-2\, b_x^2 e^4 \Delta \left(\frac{1}{3\pi}+\frac{b_x^2}{48 \pi}-4\,\Delta\right)+\mathcal{O}(e^8).
\ee
One can attempt to determine $\Delta_{\textrm{uni}}$ by imposing the trace conserving condition in \refeq{eq:Bb_tr2}. The result would depend on the degree of polarization of the initial qubits, a feature that strongly suggests that the corresponding unitary transformation does not really define a global unitary channel.

This observation is confirmed by looking explicitly at the eigenvalues of $\tilde{\rho}_{\textrm{out}}^{(++)}$, where the impact of extra coherences becomes even more transparent. At $\mathcal{O}(e^4)$, they are given by
\bea
\eta_1&=&\frac{1}{4}\left(1-b_x^2 \right)\dots\\
\eta_2&=&\frac{1}{4}\left(1-b_x^2 \right)+\frac{b_x^2 e^4 \Delta}{24\,\pi}\dots\\
\eta_3&=&\frac{1}{4}\left(1-b_x \right)^2+b_x e^4 \Delta\left(\frac{1}{3\pi}-\frac{b_x}{48 \pi}-4\Delta\right)\dots\\
\eta_4&=&\frac{1}{4}\left(1+b_x \right)^2-b_x e^4 \Delta\left(\frac{1}{3\pi}+\frac{b_x}{48\pi}-4\Delta\right)\dots
\eea
One can see that, as we increase the purity of the initial state ($b_x$ approaching~1) and move to large $\Delta<\Delta_{\textrm{uni}}$, eigenvalue~$\eta_3$ becomes negative. The reduced density matrix loses its positivity, indicating a breakdown of its physical interpretation. We reiterate that physically reasonable initial states in a scattering event imply very small values of $\Delta\ll\Delta_{\textrm{uni}}$, corresponding to wave packets highly collimated in momentum. For this reason, the theory remains well defined in all regions of physical interest. However, the eventual loss of positivity at $\Delta=\Delta_{\textrm{uni}}$ implies that a unitary channel cannot be established in Bhabha scattering following the steps of the previous sections. 

\subsection{Electron-positron annihilation into additional flavors\label{sec:QED_flav}}

One may wonder whether the inability to extract a unitarity bound on the gauge coupling that we observe in Bhabha scattering extends to more realistic settings in which an electron-positron pair can annihilate into other fermions of the Standard Model as well. In this subsection we investigate the impact of these extra flavor degrees of freedom on the properties of the scattered-state density matrix.

Let us allow for an additional annihilation channel. Consider, for example, $e^- e^+ \to \tau^- \tau^+$. One partition is added to the Hilbert space, so to include an additional qubit,
\be\label{eq:flav_Hil}
\mathcal{H}_{\textrm{tot}}=\mathcal{H}_{\textrm{mom}}\otimes \mathcal{H}_{\textrm{q}}=L^2(\mathbb{R}^3\otimes \mathbb{R}^3)\otimes\left[\mathbb{C}^{4}\otimes\mathbb{C}^4\right].
\ee
We focus here on electron-positron collisions at c.o.m.~energy $E_{\textrm{cm}}\gg m_e$. In the case of two flavors, $e$ and $\tau$, at fixed $E_{\textrm{cm}}$ we assume that $e$-type wave packets $\phi_{A,B}^{e}(\mathbf{p}_{a,b})$ are sharply peaked about $|\mathbf{p}_{A,B}|=\sqrt{E^2_{\textrm{cm}}-m_e^2}$, while $\tau$-type wave packets $\phi_{A,B}^{\tau}(\mathbf{p}_{a,b})$ are sharply peaked about $|\mathbf{p}_{A,B}|=\sqrt{E^2_{\textrm{cm}}-m_{\tau}^2}$, where, in all cases, $\mathbf{p}_A$ and $\mathbf{p}_B$ are oriented along the $z$-direction with opposite sign. Under these conditions, a sufficiently narrow width of the wave packets guarantees that the momentum-traced density matrix remains diagonal in flavor, with the exception of the important limit $E_{\textrm{cm}}\to\infty$ (we will come back to this below). This property can be easily extended to a larger number of flavors as long as they are not mass-degenerate and the respective wave packets are all sharply peaked.  

The tree-level scattering amplitudes can be derived straightforwardly for the two-flavor case and are presented explicitly in  Appendix~\ref{app:Amplitudes}. Starting with the transverse electron-positron initial state given in \refeq{eq:tra_in1}, we apply the $S$-matrix, impose optical theorem, trace out the momentum degrees of freedom, and expand to $\mathcal{O}(e^4)$. The reduced density matrix in $\mathbb{C}^4\otimes \mathbb{C}^4$ can be expressed as nested $4\times 4$ matrices. In the flavor basis \{$|ee\rangle,|e\tau\rangle,|\tau e\rangle,|\tau \tau\rangle$\} the external shell reads 
\be\label{eq:flav_mat}
\tilde{\rho}_{\textrm{out}}^{(++)}=\begin{pmatrix}
\tilde{\rho}_{ee,ee} & 0 & 0 & \tilde{\rho}_{ee,\tau\tau}\, \delta(\beta_{\tau}-1) \\
0 & 0 & 0 & 0 \\
0 & 0 & 0 & 0 \\
\tilde{\rho}_{\tau\tau,ee}\,\delta(\beta_{\tau}-1) & 0 & 0 & \tilde{\rho}_{\tau\tau,\tau\tau}
\end{pmatrix},
\ee
where we have defined $\beta_{\tau}=\sqrt{1-\frac{m_{\tau}^2}{E^2_{\textrm{cm}}}}$.

Several elements in \refeq{eq:flav_mat} are identically zero, due to the gauge interaction being flavor conserving. The delta functions are meant to enforce the requirement that off-diagonal elements be zero, with the exception of the extreme relativistic regime at $\beta_{\tau}=1$ ($E_{\textrm{cm}}\gg m_{\tau}$), where electron and tau become both effectively massless. 

At generic $b_x$ and $\beta_{\tau}$, the electron diagonal block reads, 
in the $\{|RR\rangle$, $|RL\rangle$, $|LR\rangle$, $|LL\rangle\}$ basis,
\be\label{eq:flav_trans}
\tilde{\rho}_{ee,ee}=\tilde{\rho}_{\textrm{in}}^{(++)}+\Delta\,\delta\tilde{\rho}^{(++)}_{ee},
\ee
where $\delta\tilde{\rho}^{(++)}_{ee}$ modifies  $\delta\tilde{\rho}^{(++)}$ given in \refeq{eq:corr_Bb} by the probability of scattering to the taus:
\be
\delta\tilde{\rho}^{(++)}_{ee}=b_x\, \delta\tilde{\rho}^{(++)}-\frac{\beta_{\tau}\, e^4}{8 \pi}\left(1-\frac{\beta_{\tau}^2}{3} \right)\left(\begin{array}{cccc}
0 & \frac{b_x}{2} & \frac{b_x}{2} & 0 \\
\frac{b_x}{2} & 1 & b_x^2 & \frac{b_x}{2}  \\
\frac{b_x}{2}  &  b_x^2  & 1 & \frac{b_x}{2} \\
0  & \frac{b_x}{2} & \frac{b_x}{2}  & 0
\end{array}\right)+\mathcal{O}(e^6)\,.
\ee
The tau diagonal block reads 
\be\label{eq:rhotautau}
\tilde{\rho}_{\tau\tau,\tau\tau}=\frac{\beta_{\tau}\, e^4 \Delta}{12 \pi}\left(\begin{array}{cccc}
\frac{1}{2}\left(1-\beta_{\tau}^2\right) & 0 & 0 & \frac{1}{2}\left(1-\beta_{\tau}^2\right)\\
0 & 1 & \frac{b_x^2 }{2} & 0 \\
0 &  \frac{b_x^2 }{2} & 1 & 0 \\
\frac{1}{2}\left(1-\beta_{\tau}^2\right) & 0 & 0 & \frac{1}{2}\left(1-\beta_{\tau}^2\right)
\end{array}\right)+\mathcal{O}(e^6)\,.
\ee

In the ultrarelativistic limit, $E_{\textrm{cm}}\gg m_{\tau}$, one can set $\beta_{\tau}=1$
so that the tau is effectively massless. A quantum superposition of the two flavors is thus allowed. The off-diagonal pieces in $\tilde{\rho}_{\textrm{out}}^{(++)}$ take the form, 
\begin{eqnarray}
    \tilde \rho_{ee,\tau\tau}=\frac{e^2\,\Delta}{2}\begin{pmatrix}
        0 & b_x\left(i-\frac{7 e^2}{24\pi} \right) & b_x\left(i-\frac{7 e^2}{24\pi} \right) & 0 \\
        0 & i & b_x^2 \left(i-\frac{e^2}{12\pi} \right) & 0 \\
        0 &  b_x^2 \left(i-\frac{e^2}{12\pi} \right) & i & 0  \\
        0 & b_x\left(i-\frac{7 e^2}{24\pi} \right) &  b_x\left(i-\frac{7 e^2}{24\pi} \right) & 0
    \end{pmatrix}+\mathcal{O}(e^6)\,,
\end{eqnarray}
and $\tilde \rho_{\tau\tau, ee}= \tilde \rho_{ee,\tau\tau}^\dagger$. Given Eqs.~(\ref{eq:flav_trans})-(\ref{eq:rhotautau}),
it is easy to verify that $\textrm{Tr}\,\tilde{\rho}_{\textrm{out}}^{(++)}=1$ at the desired expansion order.

Incidentally, before we proceed to tracing out the flavor partition from the density matrix, let us just reiterate that the helicity entanglement derived from Eqs.~(\ref{eq:flav_trans})-(\ref{eq:rhotautau}) has a different interpretation and different numerical value than the spin entanglement that is measured at collider experiments, which can be only computed \textit{after} one projects to one specific final state in momentum and flavor. To draw a direct 
comparison with the projection approach, we contrast in Appendix~\ref{app:proj} the concurrence obtained by projecting to the tau final state the momentum-reduced density matrix~(\ref{eq:flav_mat}) with the equivalent results of Ref.\cite{Fang:2026ddi}, which were also obtained under the assumption of an $e^- e^+$ initial state with transverse polarization. 

For a full comparison with pure Bhabha scattering, 
let us trace out the flavor degree of freedom from the density matrix. One writes 
\be
\tilde{\tilde{\rho}}_{\textrm{out}}^{(++)}=\textrm{Tr}_{\mathbb{C}^4}\tilde{\rho}_{\textrm{out}}^{(++)}=\tilde{\rho}_{ee,ee}+\tilde{\rho}_{\tau\tau,\tau\tau}\,,
\ee
where the explicit form of the matrices on the r.h.s.~is given in Eqs.~(\ref{eq:flav_trans})-(\ref{eq:rhotautau}).

At $b_x>0$ the scattering interaction further decreases the purity with respect to the initial state,
\be\label{eq:Bb_tr3}
\textrm{Tr}\,(\tilde{\tilde{\rho}}_{\textrm{out}}^{(++)})^2=\textrm{Tr}\,(\tilde{\rho}^{(++)}_{\textrm{in}})^2
-2\, b_x^2 e^4 \Delta \left(\frac{1}{3\pi}+\frac{b_x^2}{48\pi}+\frac{\beta_{\tau}\left(5+2 b_x^2\right)-\beta_{\tau}^3\left(1+b_x^2\right)}{48 \pi}-4\,\Delta\right)\dots
\ee
As before, since these $\beta_{\tau}$-dependent terms appear at $\mathcal{O}(e^4)$, the leading concurrence remains unchanged with respect to \refeq{eq:conc_mo}, even after the opening of additional scattering channels. 

Given the reduced density matrix $\tilde{\tilde{\rho}}_{\textrm{out}}^{(++)}$, it is straightforward to show, somewhat disappointingly, that the addition of one or more final-state flavors to the scattering process does not ameliorate the positivity problem we encountered in Bhabha scattering. Rather, it makes it slightly worse. At high purity and large $\Delta$ one of the eigenvalues of the reduced density matrix becomes negative. This phenomenon prevents the establishment of a unitary channel with support on the $\mathbb{C}^4$ helicity space. 

We conclude this subsection with a comment on the longitudinal ($z$-direction) polarization of the initial state. The amplitudes collected in \refeq{eq:Bhabha_tree} and Appendix~\ref{app:Amplitudes} feature the same zeros of the off-diagonal elements at $c_{\theta}=1$ as the Moeller scattering amplitude. Thus, the same observations we presented at the end of \refsec{sec:Moeller} apply unchanged to $e^- e^+$ scattering. Starting with a longitudinally polarized initial state, the concurrence of the scattered state is zero at $\mathcal{O}(e^2)$ and a unitarity bound in the fashion of \refeq{eq:Mo_uni_bound} cannot be extracted.  

\subsection{Tracing out the helicity degrees of freedom}

As is clear from our analysis, we sort of hit a roadblock in Bhabha scattering and $e^- e^+$ annihilation more generally. It seems that it might not be possible to establish a unitary quantum channel over the helicity space.  
We concluded the previous subsection by stating that, given
any initial state with some degree of longitudinal polarization, the 
concurrence after scattering remains zero at the leading $\mathcal{O}(e^2)$ order, due to the absence of coherences in the final-state density matrix. On the other hand,
while \refeq{eq:corr_Bb} shows that the scattering of an initial state with transverse polarization will indeed generate coherences of the density matrix, the final state suffers, for certain configurations of the full Hilbert space, a positivity problem. The addition of scattering channels into more massive flavors does not seem to improve the situation. 

In the two-flavor configuration, a solution comes about by recalling that, when $E_{\textrm{cm}}\gg m_{\tau}$ (or $\beta_{\tau}=1$) the density matrix in \refeq{eq:flav_mat} can present additional coherences, due to the fact that in this limiting situation electrons and taus are seen by the system as ``massless'' particles of different flavors. We can therefore set $\beta_{\tau}=1$ and trace out the helicity degrees of freedom from $\tilde{\rho}_{\textrm{out}}^{(++)}$ to obtain a flavor qubit system in the \{$|ee\rangle,|e\tau\rangle,|\tau e\rangle,|\tau \tau\rangle$\} basis,

\be\label{eq:rhofl}
\hat{\rho}_{\textrm{out}}=\textrm{Tr}_{\mathbb{C}^4}\tilde{\rho}_{\textrm{out}}^{(++)}=\hat{\rho}_{\textrm{in}}-e^2\Delta \left(\begin{array}{cccc}
0 & 0 & 0 & -i \\
0 & 0 & 0 & 0 \\
0 & 0 & 0 & 0 \\
i & 0 & 0 & 0
\end{array}\right)-\frac{e^4\Delta}{6\pi}\left(\begin{array}{cccc}
1 & 0 & 0 & 0 \\
0 & 0 & 0 & 0 \\
0 & 0 & 0 & 0 \\
0 & 0 & 0 & -1
\end{array}\right)+\mathcal{O}(e^6)\,,
\ee
where the reduced initial state $\hat{\rho}_{\textrm{in}}=|ee\rangle \langle ee|$ is pure. Note that the information on the degree of mixedness of the full initial state in the Hilbert space~(\ref{eq:flav_Hil}) has been lost.  

The trace of the density matrix squared reads
\be
\textrm{Tr}\,\hat{\rho}_{\textrm{out}}^2=1-2 e^4\Delta\left(\frac{1}{6 \pi}-\Delta \right)+\mathcal{O}(e^8)\,,
\ee
which leads to $\Delta_{\textrm{uni}}=1/(6\pi)$. It is straightforward to verify that there exists a unitary transformation connecting the initial state $\hat{\rho}_{\textrm{in}}$ to the final state in \refeq{eq:rhofl}: $U=e^{i\Delta_{\textrm{uni}}K}$, where matrix~$K$ is given in \refeq
{eq:kmatrix}.

Having established a unitary channel, we can now proceed to derive the upper bound on the coupling constant. The concurrence reads in this case
\be
C=2 e^2 \Delta+\mathcal{O}(e^4)\,,
\ee
resulting in the perturbative unitarity bound
\be\label{eq:bound}
C(\Delta_{\textrm{uni}})\leq 1\,\, \Rightarrow \,\, e^2\leq 3 \pi\,.
\ee

As was the case in Moeller scattering, extracting perturbative unitarity bounds from the partial-wave expansion in Bhabha scattering typically requires a renormalization scheme to absorb the singularities in the forward direction of the $t$-channel, a hurdle that our approach allows us to bypass. On the other hand, we can make a quantitative comparison of the bound in \refeq{eq:bound} with the bound obtained from the leading partial-wave expansion of the off-diagonal elements in \refeq{eq:Bhabha_tree}, which stem from the $s$-channel and are therefore finite. 
One writes, 
\be\label{eq:pw_QED}
a_l(s)=\frac{1}{32\pi}\int_{-1}^1d(\cos\theta)d^l_{m,m'}(\theta)\mathcal{M}(\theta,s),
\ee
where the integrand convolves the terms regular at $c_{\theta}=1$ in \refeq{eq:Bhabha_tree}, 
$-e^2\left(1-\cos\theta\right)$, 
with the corresponding Wigner~D-function,
\be
d^1_{1,-1}(\theta)=\frac{1-\cos\theta}{2}\,.
\ee
Imposing \refeq{eq:pwave}
yields $e^2\leq 12\pi$, 
which is four times less constraining than the bound in 
\refeq{eq:bound}.

\section{Summary and conclusions}\label{sec:conc}

In this work, we investigated the properties of quantum channels in the framework of $2\to2$ relativistic scattering described by the perturbative $S$-matrix. In this picture the scattered particles are defined as vectors in the Hilbert space spanned by momentum and discrete quantum numbers regarded as qubits. Our main goal is to identify the conditions under which a quantum channel, induced by tracing the momentum degrees of freedom out of the scattered system, becomes unitary. We then used qubit entanglement to derive  perturbative unitarity bounds on the strength of interactions that mediate  a scattering event.

We reached several conclusions. First, it is not always guaranteed that a unitary channel can be established in the perturbative expansion approach. We showed that while this is the case for contact interactions of flavor qubits in the 2HDM, in QED the issue strongly depends on the type of the scattering process. Moeller scattering allows a unitary channel for a family of transversely polarized helicity states, but fails to do so with longitudinally polarized initial states.  
Conversely, in Bhabha scattering and electron-positron annihilation the unitarity of the quantum channel is inconsistent with the positivity of the density matrix at our working order in perturbation theory. On the other hand, once we extend the Hilbert space to include an additional flavor qubit, a unitary quantum channel in the flavor subsystem can be established in the ultrarelativistic regime. 
These observations could suggest that whether or not a unitary channel can be defined is closely related to symmetry properties both of the scattering amplitude and the initial two-qubit state. We leave a deeper investigation of this issue for future work.

A second conclusion we reached, is that the perturbative unitarity bounds on the interaction strength that can be extracted by imposing that the concurrence is maximal in a unitary channel are consistent with, or even stronger than the usual constraints on partial-wave amplitudes. Moreover, in the case of QED processes, we were able to overcome some of the complications undermining the traditional partial-wave expansion approach, which typically requires a renormalization scheme to reabsorb the singularities of forward scattering.

This work can be extended in several directions. It would be interesting to widen the analysis to quantum field theories that feature interactions of different types (trilinear scalar couplings, pure gauge 3-point and 4-point functions, etc.), so to identify patterns that facilitate the emergence of unitary channels in qudit subsystems. Also very interesting would be to work at higher order in the perturbative expansion and verify whether our QIT-based perturbative unitarity bounds are modified. In particular, one could explore the possibility that, if one were ultimately able to generate entanglement at a higher order in the coupling constant in Moeller scattering with longitudinally polarized beams, a unitary channel in the helicity subsystem could be established globally. More ambitiously, it appears to us that the analysis presented here, as well as in our previous work\cite{Kowalska:2025qmf}, 
would benefit from a comprehensive program investigating nonperturbative approaches to entanglement in scattering theory. Nevertheless, we hope that with this study we made another small step towards establishing a dictionary between scattering theory and the concepts of QIT.

\newpage

\begin{center}
\textbf{ACKNOWLEDGMENTS}
\end{center}
This work is supported in part by the National Science Centre (Poland) under the research Grant No.~2020/38/E/ST2/00126. SC cordially thanks Priyanka Lamba and Wojciech Kotlarski for helpful discussions. KK and EMS are grateful to the CERN Theoretical Physics Department for the kind hospitality during the completion of this paper. 
\bigskip

\appendix
\addcontentsline{toc}{section}{Appendices}

\section{Amplitude matrices in QED\label{app:Amplitudes}}

In this appendix, we recall the tree-level amplitudes for electron-positron annihilation onto two distinct flavors: $e^- e^+ \to e^- e^+, \tau^- \tau^+$. All amplitudes are computed following the Jacob-Wick convention\cite{Jacob:1959at} for their phases.
One writes
\begin{equation}
    \mathcal{M}_{(\textrm{tree})}=\begin{pmatrix}
        \mathcal{M}_{(\textrm{tree})}^{ee,ee} & \mathcal{M}_{(\textrm{tree})}^{ee,\tau\tau} \\
        \mathcal{M}_{(\textrm{tree})}^{\tau\tau,ee} & \mathcal{M}_{(\textrm{tree})}^{\tau\tau,\tau\tau}
    \end{pmatrix},
\end{equation}
where $\mathcal{M}_{(\textrm{tree})}^{ee,ee}$ was presented in \refeq{eq:Bhabha_tree}.

The three  remaining amplitudes are the following. First $\tau^- \tau^-\to e^- e^+$,
\begin{multline}
\mathcal{M}^{ee,\tau\tau}_{(\textrm{tree})}=\\
e^2\begin{pmatrix}
0 & 0 & 0 & 0\\ 
-e^{-i\phi}\sqrt{\left(1-c_{\theta}^2\right)\left(1-\beta_{\tau}^2\right)} & -\left(1+c_{\theta}\right) & -\left(1-c_{\theta}\right)e^{-2 i \phi} & -e^{-i\phi}\sqrt{\left(1-c_{\theta}^2\right)\left(1-\beta_{\tau}^2\right)}  \\
e^{i\phi}\sqrt{\left(1-c_{\theta}^2\right)\left(1-\beta_{\tau}^2\right)} &   -\left(1-c_{\theta}\right)e^{2 i \phi} & -\left(1+c_{\theta}\right) & e^{i\phi}\sqrt{\left(1-c_{\theta}^2\right)\left(1-\beta_{\tau}^2\right)} \\
0 & 0 & 0 & 0
\end{pmatrix},
\end{multline}
where $c_{\theta}\equiv \cos\theta$ and we have defined $\beta_{\tau}=\sqrt{1-\frac{m_{\tau}^2}{E^2_{\textrm{cm}}}}$.

Next, $e^- e^+ \to \tau^- \tau^+$,
\be
\mathcal{M}^{\tau\tau,ee}_{(\textrm{tree})}
=e^2\begin{pmatrix}
0 & e^{i\phi}\sqrt{\left(1-c_{\theta}^2\right)\left(1-\beta_{\tau}^2\right)} & -e^{-i\phi}\sqrt{\left(1-c_{\theta}^2\right)\left(1-\beta_{\tau}^2\right)} & 0\\ 
0 & -\left(1+c_{\theta}\right) & -\left(1-c_{\theta}\right)e^{-2 i \phi} & 0  \\
0 & -\left(1-c_{\theta}\right)e^{2 i \phi} & -\left(1+c_{\theta}\right) & 0  \\
0 & e^{i\phi}\sqrt{\left(1-c_{\theta}^2\right)\left(1-\beta_{\tau}^2\right)} & -e^{-i\phi}\sqrt{\left(1-c_{\theta}^2\right)\left(1-\beta_{\tau}^2\right)} & 0
\end{pmatrix}.
\ee

Finally, $\tau^- \tau^+ \to \tau^- \tau^+$,
\begin{eqnarray}
    \mathcal{M}^{\tau\tau,\tau\tau}_{(\textrm{tree})}
    =\frac{e^2}{\beta_{\tau}^2\left(1-c_{\theta}\right)}\begin{pmatrix}
    A_1 & -A_2 e^{i\phi} & A_2 e^{-i\phi} & -A_3 \\[6pt]
    A_2 e^{-i\phi} & A_4 & -A_5 e^{-2i\phi} & A_2 e^{-i\phi} \\[6pt]
    -A_2 e^{i\phi} & -A_5 e^{2i\phi} & A_4 & -A_2 e^{i\phi} \\[6pt]
    -A_3 & -A_2 e^{i\phi} & A_2 e^{-i\phi} & A_1
    \end{pmatrix},
\end{eqnarray}
where    
\begin{align}
A_1 &\equiv 1+c_{\theta}+\left(3-2 c_{\theta}+c_{\theta}^2\right)\beta_{\tau}^2-\left(c_{\theta}-1\right)c_{\theta}\beta_{\tau}^4 \\
A_2 &\equiv \sqrt{\left(1-c_{\theta}^2\right)\left(1-\beta_{\tau}^2\right)}\left(1-\beta_{\tau}^2+c_{\theta}\beta_{\tau}^2\right) \\
A_3 &\equiv \left(1-\beta_{\tau}^2 \right)\left(1+c_{\theta}\beta_{\tau}^2 \right)\left(1-c_{\theta}\right)\\
A_4 &\equiv \left(1+c_{\theta}\right)\left(1+c_{\theta}\beta_{\tau}^2 \right)\\
A_5 &\equiv \left(c_{\theta}-1\right)\left(1-2\beta_{\tau}^2+c_{\theta}\beta_{\tau}^2\right).
\end{align}

\section{Comparison with the projection approach\label{app:proj}}

In this appendix, we draw a direct comparison between the concurrence 
obtained by projecting the momentum-reduced density matrix~(\ref{eq:flav_mat}) to the tau flavor, and the concurrence computed in Ref.\cite{Fang:2026ddi}, which was obtained by projecting the final-state density matrix to the tau flavor while maintaining the explicit angular dependence of the momentum partition. While both approaches considered in this appendix effectively project to one specific flavor, the fundamental difference arises from the fact that, in our formalism, the reduced density matrix of the scattered state does not have access to information regarding the momentum angular distribution.

We remind the reader that in the projection approach thus defined\cite{Cervera-Lierta:2017tdt,Fonseca:2021uhd,Fedida:2022izl, Blasone:2024dud,Blasone:2024jzv,Blasone:2025ddi, Fang:2026ddi} the qubit density matrix is generated by projecting the scattering outcome matrix~$\rho_{\textrm{out}}$ onto a fixed momentum vector $|\mathbf{p}_C\mathbf{p}_D\rangle$. Schematically, the final state is given by 
\be\label{eq:rhopr}
\tilde{\rho}_{\textrm{proj}}(\mathbf{p}_C,\mathbf{p}_D)=\frac{\mathcal{M}(\mathbf{p}_C,\mathbf{p}_D)\tilde{\rho}_{\textrm{in}}\mathcal{M}(\mathbf{p}_C,\mathbf{p}_D)^\dagger}{\tr\left[\mathcal{M}(\mathbf{p}_C,\mathbf{p}_D)\tilde{\rho}_{\textrm{in}}\mathcal{M}(\mathbf{p}_C,\mathbf{p}_D)^\dagger\right]}\,,
\ee
which carries the full angular dependence of $|\mathbf{p}_C\mathbf{p}_D\rangle$, expressed through the tree-level amplitude $\mathcal{M}(\mathbf{p}_C,\mathbf{p}_D)$.
Conversely, in our approach 3-momentum is traced out of the final-state matrix after imposing unitarity at the appropriate order via the optical theorem. 

Let us consider the scattering process $e^- e^+\to \tau^- \tau^+$, where the initial state $\tilde{\rho}_{\textrm{in}}$ is partially polarized in the $x$-direction like in \refeq{eq:tra_in}. We can project to the $ \tau^- \tau^+$ final state by normalizing \refeq{eq:rhotautau}. One finds,
\be\label{eq:rhotautauN}
\tilde{\rho}_{\tau\tau,\textrm{N}}=\left(\begin{array}{cccc}
\frac{1}{2}-\frac{1}{3-\beta_{\tau}^2} & 0 & 0 &\frac{1}{2}-\frac{1}{3-\beta_{\tau}^2} \\
0 &\frac{1}{3-\beta_{\tau}^2}  & \frac{b_x^2}{6-2\beta_{\tau}^2} & 0\\
0 & \frac{b_x^2}{6-2\beta_{\tau}^2}   & \frac{1}{3-\beta_{\tau}^2} & 0 \\
\frac{1}{2}-\frac{1}{3-\beta_{\tau}^2} & 0 & 0  & \frac{1}{2}-\frac{1}{3-\beta_{\tau}^2}
\end{array}\right).
\ee
Note that $e^2$ and $\Delta$ have disappeared from the normalized state~$\tilde{\rho}_{\tau\tau,\textrm{N}}$. This makes physical sense: once a scattered state is measured, both the strength of the interaction and the degree of overlap between the initial and final wave packets are not 
relevant to the quantum correlations between the discrete degrees of freedom. On the other hand, 
the scattered state conserves memory of the degree of coherence of the initial state, parametrized by $b_x$, and it also depends on how relativistic is the scattering process, parametrized by $\beta_{\tau}$. Increasing
$\beta_{\tau}$ reduces the impact of the tau mass and thus increases the purity of the final state, making it more likely that the qubits be entangled. 

To this regard, the purity of the final state 
in \refeq{eq:rhotautauN} is given by 
\be
\textrm{Tr}\, \tilde{\rho}_{\tau\tau,\textrm{N}}^2=\frac{6+b_x^4-4 \beta_{\tau}^2+2 \beta_{\tau}^4}{2\left( 3-\beta_{\tau}^2\right)^2}<1\,,
\ee
for all $0\leq b_x,\beta_{\tau}\leq 1$. The concurence reads
\be\label{eq:conc_proj}
C=\max\left[0,-\frac{1-b_x^2-\beta_{\tau}^2}{3-\beta_{\tau}^2}\right].
\ee

 \begin{figure}[t]
	\centering%
		\includegraphics[width=0.5\textwidth]{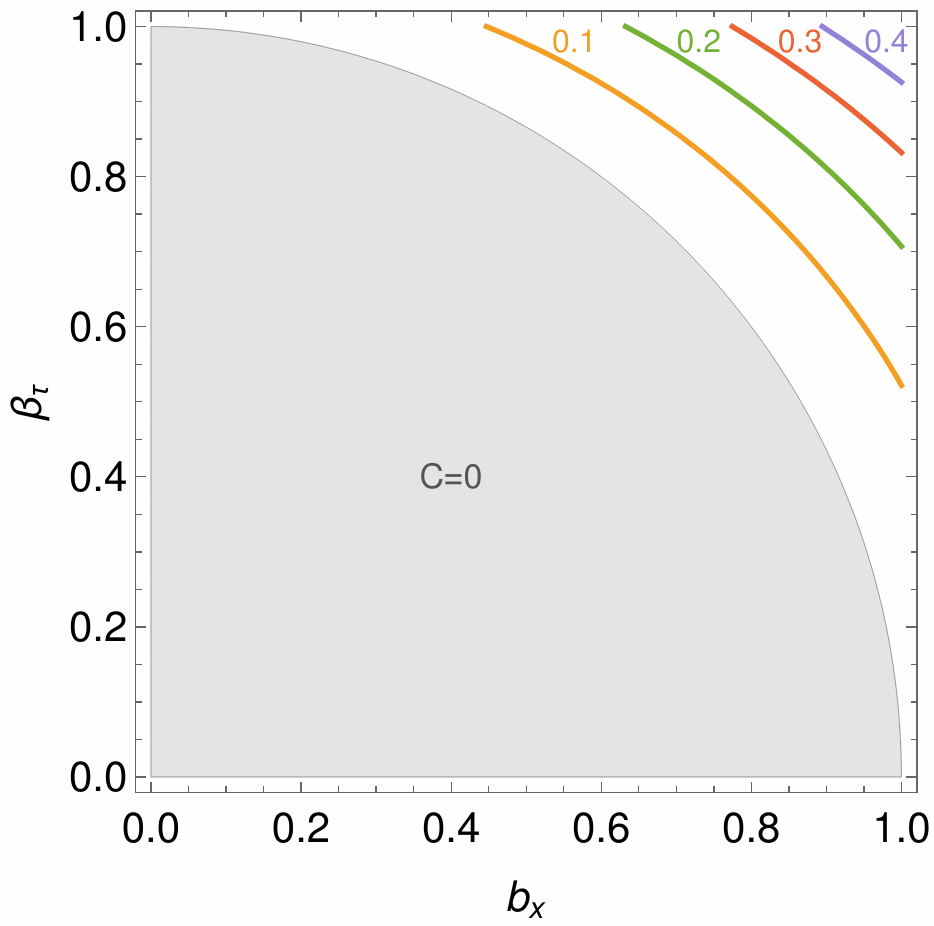}
\caption{Isocontours of the concurrence of the $\tau^{-}\tau^{+}$ final state in the~($b_x$, $\beta_{\tau}$) plane.}
\label{fig:proj1}
\end{figure}

In \reffig{fig:proj1} we present isocontours of the concurrence~(\ref{eq:conc_proj}) in the~($b_x$, $\beta_{\tau}$) plane. For a direct comparison with the projection approach, the figure should be confronted, e.g., with 
Fig.~2 of Ref.\cite{Fang:2026ddi}.

One can spot similarities and differences of the two approaches right away. As we have mentioned above, in \reffig{fig:proj1}, as well as in Fig.~2 of Ref.\cite{Fang:2026ddi}, the concurrence is maximal either when the coherence of the initial state is maximal~($b_x=1$) or at ultrarelativistic energies~($\beta_{\tau}=1$). Unlike in Fig.~2 of Ref.\cite{Fang:2026ddi}, however, no Bell state can be ever generated when the momentum is traced out and 
the final state remains mixed over the range $0\leq b_x,\beta_{\tau}\leq 1$. Moreover, there appears to remain in \reffig{fig:proj1} a large region of the parameter space, indicated in gray, for which the final state cannot even be entangled. This significant discrepancy with Ref.\cite{Fang:2026ddi} is a consequence of the QED amplitude carrying non-trivial phases of the azimuthal angle~$\phi$. When the information on the angular distribution is not available (i.e., 3-momentum is traced out), the phases have to be averaged over, with the inevitable consequence that the final state loses some of its coherence with respect to the formulation in \refeq{eq:rhopr}, which reproduces explicitly the angular distribution of the amplitude.

\bibliographystyle{utphysmcite}
\bibliography{mybib}

\end{document}

%% file: macros.tex
\newcommand{\newc}{\newcommand*}

\long\def\begincomment#1\endcomment{%
        \begingroup\sf\baselineskip12pt#1\endgroup}

\newc{\etal}{\textrm{et al.}} 
\newc{\eg}{\textrm{e.g.}} 
\newc{\ie}{\textrm{i.e.}}
\newc{\etc}{\textrm{etc}}
\newc\vs{\textrm{vs.}}
\newc{\cl}{\rm {C.L.}}

\newc{\ev}{\ensuremath{\,\mathrm{eV}}}
\newc{\kev}{\ensuremath{\,\mathrm{keV}}}
\newc{\mev}{\ensuremath{\,\mathrm{MeV}}}
\newc{\gev}{\ensuremath{\,\mathrm{GeV}}}
\newc{\tev}{\ensuremath{\,\mathrm{TeV}}}
\newc{\MeV}{\mev} 
\newc{\TeV}{\tev}
\newc{\invpb}{\ensuremath{/\text{pb}}}
\newc{\invfb}{\ensuremath{\,\textrm{fb}^{-1}}}
\newc\nb{\ensuremath{\,\mathrm{nb}}} \newc\pb{\ensuremath{\,\mathrm{pb}}} \newc\fb{\ensuremath{\,\mathrm{fb}}}
\newc\pc{\ensuremath{\,\mathrm{pc}}}
\newc\kpc{\ensuremath{\,\mathrm{kpc}}}
\newc\mpc{\ensuremath{\,\mathrm{Mpc}}}
\newc\ps{\ensuremath{\,\mathrm{ps}}} 
\newc\cmeter{\ensuremath{\,\mathrm{cm}}} 
\newc\meter{\ensuremath{\,\mathrm{m}}} 
\newc\kmeter{\ensuremath{\,\mathrm{km}}}
\newc\second{\ensuremath{\,\mathrm{s}}}
\newc\msecond{\ensuremath{\,\mathrm{ms}}}
\newc\nsecond{\ensuremath{\,\mathrm{ns}}}
\newc\psecond{\ensuremath{\,\mathrm{ps}}}

\newc{\chisqmin}{\ensuremath{\chi^2_{\mathrm{min}}}}
\newc{\Delchisq}{\ensuremath{\Delta\chi^2}}
\newc{\chisq}{\ensuremath{\chi^2}}
\newc{\like}{\ensuremath{\mathcal{L}}}

\newc\lsim{\ensuremath{\mathrel{\rlap{\lower4pt\hbox{\hskip1pt$\sim$}}\raise1pt\hbox{$<$}}}}
\newc\gsim{\ensuremath{\mathrel{\rlap{\lower4pt\hbox{\hskip1pt$\sim$}}\raise1pt\hbox{$>$}}}}
\newc{\VEV}[1]{\ensuremath{\langle #1 \rangle}}
\newc{\dl}{\ensuremath{\stackrel{\leftarrow}{D}}}
\newc{\dr}{\ensuremath{\stackrel{\rightarrow}{D}}}

\newc{\bcenter}{\begin{center}}    \newc{\ecenter}{\end{center}}
\newc{\bfl}{\begin{flushleft}}    \newc{\efl}{\end{flushleft}}
\newc{\bfr}{\begin{flushright}}    \newc{\efr}{\end{flushright}}

\newc{\bi}{\begin{itemize}}
\newc{\ei}{\end{itemize}}
\newc{\bed}{\begin{description}}
\newc{\eed}{\end{description}}
\newc{\ben}{\begin{enumerate}}
\newc{\een}{\end{enumerate}}

\newc{\be}{\begin{equation}}
\newc{\ee}{\end{equation}}
\newc{\bea}{\begin{eqnarray}}
\newc{\eea}{\end{eqnarray}}
\newc{\bfle}{\begin{flalign}}
\newc{\efle}{\end{flalign}}
\newc{\ra}{\rightarrow}

\newc{\alphas}{\ensuremath{\alpha_s}}
\newc{\alphatwo}{\ensuremath{\alpha_2}}
\newc{\alphaone}{\ensuremath{\alpha_1}}
\newc{\alphai}[1]{\ensuremath{\alpha_{#1}}}
\newc{\alphaem}{\ensuremath{\alpha_{\mathrm{em}}}}
\newc{\alphaeff}{\ensuremath{\alpha_{\mathrm{eff}}}}
\newc{\sineff}{\ensuremath{\sin \theta_{\mathrm{eff}}}}
\newc{\sinsqeff}{\ensuremath{\sin^2 \theta_{\mathrm{eff}}}}
\newc{\dalphahad}{\ensuremath{\Delta \alpha_{\mathrm{had}}}}
\newc{\yt}{\ensuremath{h_t}} \newc{\yb}{\ensuremath{h_b}} \newc{\ytau}{\ensuremath{h_{\tau}}}
\newc\mz{\ensuremath{M_Z}} 
\newc\mw{\ensuremath{m_W}}
\newc\mZ{\mz}        \newc\mW{\mw}
\newc\mhsm{\ensuremath{ m_{H_{\mathrm{SM}}}}}
\newc{\mtop}{\ensuremath{ m_t}}               \newc{\mtpole}{\ensuremath{ M_t}}
\newc{\mbottom}{\ensuremath{ m_b}} 
\newc{\mtau}{\ensuremath{ m_{\tau}}}
\newc{\mt}{\mtpole}
\newc{\mb}{\mbottom} 
\newc{\rtwogg}{\ensuremath{R_{h_2}(\gamma\gamma)}}
\newc{\rtwozz}{\ensuremath{R_{h_2}(ZZ)}}
\newc{\ronegg}{\ensuremath{R_{h_1}(\gamma\gamma)}}
\newc{\ronezz}{\ensuremath{R_{h_1}(ZZ)}}
\newc{\rsiggg}{\ensuremath{R_{h_\textrm{sig}}(\gamma\gamma)}}
\newc{\rsigzz}{\ensuremath{R_{h_\textrm{sig}}(ZZ)}}
\newc{\llbar}{\ensuremath{\ell\bar{\ell}}}
\newc{\tauptaum}{\ensuremath{ \tau^+\tau^-}}
\newc{\qqbar}{\ensuremath{ q\bar{q}}} \newc{\ppbar}{\ensuremath{ p\bar{p}}}
\newc{\bbbar}{\ensuremath{ b\bar{b}}} \newc{\ttbar}{\ensuremath{ t\bar{t}}}
\newc{\ffbar}{\ensuremath{ f\bar{f}}} \newc{\tautaubar}{\ensuremath{ \tau\bar{\tau}}}

\newc{\mchi}{\ensuremath{m_\neutone}}
\newc{\squark}{\ensuremath{\tilde{q}}}
\newc{\slepton}{\ensuremath{\tilde{l}}}
\newc{\gluino}{\ensuremath{\tilde{g}}} 
\newc{\mgluino}{\ensuremath{{m_{\gluino}}}}
\newc{\wino}{\ensuremath{\tilde{W}}} 
\newc{\mwino}{\ensuremath{{m_{\wino}}}}
\newc{\tone}{\ensuremath{{\tilde{t}_1}}}
\newc{\bone}{\ensuremath{{\tilde{b}_1}}}
\newc{\Hone}{\ensuremath{{\tilde{H}_{1}}}}
\newc{\Htwo}{\ensuremath{{\tilde{H}_{2}}}}
\newc{\Hhtwo}{\ensuremath{{H_{2}}}}
\newc{\qli}{\ensuremath{{\tilde{Q}_{i}}}}
\newc{\uri}{\ensuremath{{\tilde{u}_{i}}}}
\newc{\dri}{\ensuremath{{\tilde{d}_{i}}}}
\newc{\lli}{\ensuremath{{\tilde{L}_{i}}}}
\newc{\eri}{\ensuremath{{\tilde{e}_{i}}}}

\newc{\sthw}{\ensuremath{ \sin\theta_W}}              \newc{\cthw}{\ensuremath{\cos\theta_W}}
\newc{\tanthw}{\ensuremath{ \tan\theta_W}}              \newc{\cotthw}{\ensuremath{\cot\theta_W}}
\newc{\ssqthw}{\ensuremath{\sin^2 \theta_W}}
\newc{\msbar}{\ensuremath{\overline{MS}}} \newc{\drbar}{\ensuremath{\overline{DR}}}
\newc{\mtmtsmmsbar}{\ensuremath{ m_t(m_t)^{\msbar}_{{\mathrm{SM}}}}}
\newc{\mtmtsmdrbar}{\ensuremath{ m_t(m_t)^{\drbar}_{{\mathrm{SM}}}}}
\newc{\mtmtmssmdrbar}{\ensuremath{ m_t(m_t)^{\drbar}_{{\mathrm{SUSY}}}}}
\newc{\mbmbmsbar}{\ensuremath{ m_b(m_b)^{\msbar} }}
\newc{\mbmbsmmsbar}{\ensuremath{ m_b(m_b)^{\msbar}_{{\mathrm{SM}}}}}
\newc{\mbmzsmmsbar}{\ensuremath{ m_b(\mz)^{\msbar}_{{\mathrm{SM}}}}}
\newc{\mbmzsmdrbar}{\ensuremath{ m_b(\mz)^{\drbar}_{{\mathrm{SM}}}}}
\newc{\mbmzmssmdrbar}{\ensuremath{ m_b(\mz)^{\drbar}_{{\mathrm{SUSY}}}}}
\newc{\mtaumzsmmsbar}{\ensuremath{ m_{\tau}(\mz)^{\msbar}_{{\mathrm{SM}}}}}
\newc{\mtaumzsmdrbar}{\ensuremath{ m_{\tau}(\mz)^{\drbar}_{{\mathrm{SM}}}}}
\newc{\mtaumzmssmdrbar}{\ensuremath{ m_{\tau}(\mz)^{\drbar}_{{\mathrm{SUSY}}}}}
\newc{\alphasmzms}{\ensuremath{\alpha_s(M_Z)^{\overline{MS}}}}
\newc{\alphaimzms}[1]{\ensuremath{\alpha_{#1}(M_Z)^{\overline{MS}}}}
\newc{\alphaemmz}{\ensuremath{\alpha_{\mathrm{em}}(M_Z)^{\overline{MS}}}}

\newc{\mzero}{\ensuremath{{m_0}}}
\newc{\mhalf}{\ensuremath{ m_{1/2}}}
\newc{\tanb}{\ensuremath{\tan\beta}}
\newc{\azero}{\ensuremath{ A_0}}
\newc{\signmu}{\ensuremath{\rm{sgn}\,\mu}}
\newc{\atau}{\ensuremath{{A_{\tau}}}}
\newc{\mueff}{\ensuremath{\mu_{\rm{eff}}}}
\newc{\lam}{\ensuremath{{\lambda}}}
\newc{\kap}{\ensuremath{{\kappa}}}
\newc{\alam}{\ensuremath{{A_{\lambda}}}}
\newc{\akap}{\ensuremath{{A_{\kappa}}}}
\newc{\hs}{\ensuremath{ H_s}}      
\newc{\mhs}{\ensuremath{ m_{H_s}}} 
\newc{\mgut}{\ensuremath{ M_{\rm GUT}}}
\newc{\gut}{\ensuremath{{\rm GUT}}}
\newc{\mplanck}{\ensuremath{ M_{\rm P}}}      \newc{\mpl}{\ensuremath{ M_{\rm Pl}}}
\newc{\msusy}{\ensuremath{ M_{\rm SUSY}}}      \newc{\ms}{\ensuremath{ M_{\rm S}}}
 \newc{\hu}{\ensuremath{ H_u}}       \newc{\hd}{\ensuremath{ H_d}}
 \newc{\mhu}{\ensuremath{ m_{H_u}}}       \newc{\mhd}{\ensuremath{ m_{H_d}}}
 \newc{\mhuew}{\ensuremath{ m^{\ast}_{H_u}}}       \newc{\mhdew}{\ensuremath{ m^{\ast}_{H_d}}}
 \newc{\mhuewsq}{\ensuremath{ m^{\ast\, 2}_{H_u}}}       \newc{\mhdewsq}{\ensuremath{ m^{\ast\, 2}_{H_d}}}
 \newc{\mhl}{\ensuremath{m_\hl}} 
 \newc{\mhone}{\ensuremath{m_{h_1}}} 
 \newc{\mhtwo}{\ensuremath{m_{h_2}}} 
 \newc{\mhi}{\ensuremath{m_{\tilde{h}}}} 
 \newc{\mul}{\ensuremath{m_{\tilde{u}_L}}} 
 \newc{\mbone}{\ensuremath{m_{\tilde{b}_1}}}  
 \newc{\mtone}{\ensuremath{m_{\tilde{t}_1}}} 
 \newc{\ma}{\ensuremath{m_A}} 
 \newc{\mH}{\ensuremath{m_H}} 
 \newc{\maone}{\ensuremath{m_{a_1}}} 
 \newc{\matwo}{\ensuremath{m_{a_2}}}
 \newc{\hone}{\ensuremath{h_1}}
 \newc{\htwo}{\ensuremath{h_2}}
 \newc{\aone}{\ensuremath{a_1}}
 \newc{\atwo}{\ensuremath{a_2}}
 \newc{\mqthree}{\ensuremath{m_{\tilde{Q}_3}^2}}
 \newc{\muthree}{\ensuremath{m_{\tilde{u}_3}^2}}
 \newc{\mqli}{\ensuremath{m_{\tilde{Q}_{i}}}}
 \newc{\muri}{\ensuremath{m_{\tilde{u}_{i}}}}
 \newc{\mdri}{\ensuremath{m_{\tilde{d}_{i}}}}
 \newc{\mlli}{\ensuremath{m_{\tilde{L}_{i}}}}
 \newc{\meri}{\ensuremath{m_{\tilde{e}_{i}}}}
 \newc{\ts}{\ensuremath{T_{SUSY}}}

\newc{\sigsip}{\ensuremath{\sigma^{\rm SI}_{p}}}	\newc{\sigsin}{\ensuremath{\sigma^{\rm SI}_{n}}}
\newc{\sigsdp}{\ensuremath{\sigma^{\rm SD}_{p}}}	\newc{\sigsdn}{\ensuremath{\sigma^{\rm SD}_{n}}}
\newc{\sigsi}{\ensuremath{\sigma^{\rm SI}}}	\newc{\sigsd}{\ensuremath{\sigma^{\rm SD}}}
\newc{\abund}{\ensuremath{ \Omega h^2}}
\newc{\omegadm}{\ensuremath{ \Omega_{{\rm DM}}}}     \newc{\abunddm}{\ensuremath{ \Omega_{{\rm DM}} h^2}} 
\newc{\omegam}{\ensuremath{ \Omega_{{\rm m}}}}       \newc{\abundm}{\ensuremath{ \Omega_{{\rm m}} h^2}}
\newc{\omegab}{\ensuremath{ \Omega_{{\rm b}}}}	\newc{\abundb}{\ensuremath{ \Omega_{{\rm b}} h^2}}
\newc{\omegatot}{\ensuremath{ \Omega_{{\rm TOT}}}}
\newc{\omegacdm}{\ensuremath{ \Omega_{{\rm CDM}}}}   \newc{\abundcdm}{\ensuremath{ \Omega_{{\rm CDM}} h^2}}
\newc{\omegalambda}{\ensuremath{ \Omega_{\Lambda}}} \newc{\abundlambda}{\ensuremath{ \Omega_{\Lambda} h^2}}
\newc{\omegarad}{\ensuremath{ \Omega_{{\rm rad}}}}  \newc{\abundrad}{\ensuremath{ \Omega_{{\rm rad}} h^2}}
\newc{\rhocrit}{\ensuremath{ \rho_{\rm crit}}}
\newc{\rhochi}{\ensuremath{ \rho_{\chi}}}
\newc{\abunchi}{\ensuremath{\Omega_\chi h^2}}
\newc{\abundlsp}{\ensuremath{\Omega_{\rm LSP}h^2}}

\newc{\amu}{\ensuremath{ a_{\mu}}}        \newc{\amususy}{\ensuremath{ a_{\mu}^{\mathrm{SUSY}}}}
\newc{\amuexpt}{\ensuremath{ a_{\mu}^{\mathrm{expt}}}}        \newc{\amusm}{\ensuremath{ a_{\mu}^{\mathrm{SM}}}}
\newc\deltaamu{\ensuremath{\Delta a_{\mu}}} \newc{\deltaamususy}{\ensuremath{\delta a_{\mu}^{\mathrm{SUSY}}}}
\newc\gmtwo{\ensuremath{ (g-2)_{\mu}}} 
\newc{\deltagmtwomususy}{\ensuremath{\delta\left(g-2\right)_{\mu}^{\mathrm{SUSY}}}}
\newc{\deltagmtwomu}{\ensuremath{\delta\left(g-2\right)_{\mu}}}
\newc\BR{\ensuremath{\rm BR}}
\newc\bsgamma{\ensuremath{ b\rightarrow s \gamma }}
\newc\bxsgamma{\ensuremath{\overline{B}\rightarrow X_{s}\gamma}}
\newc\brbsgamma{\ensuremath{\BR\left(\bsgamma\right)}}
\newc\brbxsgamma{\ensuremath{\BR\left(\bxsgamma\right)}}
\newc\bsmumu{\ensuremath{B_s\to\mu^+\mu^-}}
\newc\brbsmumu{\ensuremath{\BR\left(B_s\to\mu^+\mu^-\right)}}
\newc\bdmmumu{\ensuremath{\overline{B}_d\to\mu^+\mu^-}}
\newc\bbbarmix{\ensuremath{\overline{B}_s\mbox{-}B_s}}      
\newc\delmbs{\ensuremath{\Delta M_{B_s}}}
\newc{\butaunu}{\ensuremath{B_u \rightarrow \tau \nu}}
\newc{\brbutaunu}{\ensuremath{\BR\left(B_u \rightarrow \tau \nu\right)}}

\newcommand*{\reffig}[1]{Fig.~\ref{#1}}
 
        \newcommand*{\refeq}[1]{Eq.~(\ref{#1})}
     \newcommand*{\refsec}[1]{Sec.~\ref{#1}}

\newcommand*{\neutone}{\ensuremath{\chi^0_1}}

\let\oldcite\cite
\renewcommand*{\cite}{~\oldcite}

\newcommand*{\hl}{\ensuremath{h}}

\newcommand*{\aket}{\ensuremath{|A\rangle}}

\newcommand*{\bbra}{\ensuremath{\langle B|}}

\newcommand*{\mfw}{\ensuremath{\mathcal{M}_{\textrm{fw}}}}

\newcommand*{\tr}{\ensuremath{\textrm{Tr}}}